\documentclass[useAMS,usenatbib]{mn2e}
\usepackage{bethmacros}
\usepackage{graphicx,subfigure}
\usepackage{amssymb}

\def\spose#1{\hbox to 0pt{#1\hss}}
\def\lta{\mathrel{\spose{\lower 3pt\hbox{$\mathchar"218$}}
     \raise 2.0pt\hbox{$\mathchar"13C$}}}
\def\gta{\mathrel{\spose{\lower 3pt\hbox{$\mathchar"218$}}
     \raise 2.0pt\hbox{$\mathchar"13E$}}}

\newcommand{\sech}{\mathrm{sech} \,}

\title[A Thick Disk from Stellar Feedback ]{MaGICC Thick Disk I: Comparing a Simulated Disk Formed with Stellar Feedback to the Milky Way}

\author[Stinson et al.]{G.\,S. Stinson$^{1}$\thanks{Email: stinson `at' mpia.de}, J. Bovy$^{2,7}$, H-W Rix$^1$, C. Brook $^3$, R. Ro\v{s}kar$^4$,
\newauthor{J. J. Dalcanton$^5$, A. V. Macci\`o$^1$,  J. Wadsley$^{6}$, H. M. P. Couchman$^6$, T. R. Quinn$^{5}$}
\vspace*{6pt}\\
$^{1}$Max-Planck-Institut f\"ur Astronomie, K\"onigstuhl 17, 69117, Heidelberg, Germany\\
$^2$Institute for Advanced Study, Einstein Drive, Princeton, NJ 08540, USA\\
$^{3}$Departamento de F\'{i}s\'{i}ca Te\'{o}rica, Universidad Aut\'{o}noma de Madrid, E-28049 Cantoblanco, Madrid, Spain\\
$^4$Institut f\"{u}r Theoretische Physik, Universit\"{a}t Z\"{u}rich, Switzerland\\
$^{5}$Astronomy Department, University of Washington, Box 351580, Seattle, WA, 98195-1580\\
$^{6}$Department of Physics and Astronomy, McMaster University, Hamilton, Ontario, L8S 4M1, Canada\\
$^7$Hubble Fellow}

\begin{document}
\maketitle
\label{firstpage}

\begin{abstract}
We analyse the structure and chemical enrichment of a Milky Way-like galaxy
with a stellar mass of $2\times10^{10}$ M$_\odot$, formed 
in a cosmological hydrodynamical simulation.  It is disk-dominated with a flat rotation curve, 
and has a disk scale length similar to the Milky Way's, but a velocity dispersion
that is $\sim$50\% higher. Examining stars in narrow [Fe/H] and [$\alpha$/Fe] 
abundance ranges, we find remarkable qualitative agreement between 
this simulation and observations: a) 
The old stars lie in a thickened distribution with a short scale length, while 
the young stars form a thinner disk, with scale lengths decreasing, as [Fe/H] increases. 
b) Consequently, there is a distinct outward metallicity 
gradient. c) Mono-abundance populations exist with a continuous distribution 
of scale heights (from thin to thick). However, the simulated galaxy has a 
distinct and substantive very thick disk ($h_z\sim1.5$ kpc), not seen in the Milky Way.
The broad agreement between simulations and observations allows us to test the 
validity of observational proxies used in the literature: we find in the 
simulation that mono-abundance populations are good proxies for single 
age populations ($<$1 Gyr) for most abundances.

\end{abstract}


\section{Introduction}
Ever since the discovery that the luminosity distribution of edge-on S0 galaxies could be well fit by including a thick and thin disk component \citep{Burstein1979}, evidence has accumulated for the ubiquity of thick disk components in galaxies.  \citet{vanderKruit1981} found that the disk of the late type galaxy NGC 891 required three separate components to fit its vertical profile, a bulge, thick and thin disk.  Number counts of dwarf stars in the Solar Neighborhood found that the Milky Way disk also was well fit using two distinct scale heights \citep{Gilmore1983}.  Subsequent observations found that every observed disk galaxy are characterized by more than one single vertical scale height, but are well fit by two components \citep{vanderKruit1982,Dalcanton2000,Dalcanton2002, Seth2005, Yoachim2006, deJong2007}.  Recently, large surveys have allowed more extensive examinations of the Milky Way disk structure. With the Sloan Digital Sky Survey, \citet{Juric2008} confirmed that the spatial structure of the Galactic disk can be well described with two distinct scale heights.  

In addition to the density structure of the stellar disks, observations have also examined the chemical abundances of the stars that make up different sub-populations in the Milky Way.  A number of spectroscopic studies have found distinct stellar abundances in stars that are kinematically associated with the thick disk \citep{Majewski1993,Gilmore1995,Fuhrmann1998,Chiba2000,Prochaska2000, Bensby2003, Bensby2005,Wyse2006,Reddy2006,Fuhrmann2008,Ruchti2010}.  Stars in the thick disk components are more $\alpha$-enhanced than kinematically cooler thin disk stars.  It is reasonable to assume that $\alpha$-enhanced stars are older since they formed from gas not yet polluted with the iron rich ejecta of longer-lived Type Ia supernovae.  

Spectroscopic surveys have begun to assemble more extensive samples and have thus allowed more detailed comparisons between stellar abundance, disk structure, and kinematics.  The Geneva-Copenhagen Survey found the largest radial [Fe/H] gradient in their youngest age group ($< 1.5$ Gyr), and progressively less gradient in their older age groups \citep[see Figure 29 of][]{Nordstrom2004}. The Sloan Digital Sky Survey found that there is a smooth transition from the metallicity of thick to thin disk stars \citep{Ivezic2008}, which is also noted in spectroscopic studies \citep{Norris1987,Haywood2013,Kordopatis2013}.  \citet{Navarro2011} used a compilation of stellar spectra from the literature to find a separation between kinematically distinct thick and thin disk populations when selected based on their position in the [$\alpha$/Fe]-[Fe/H] plane.

One of the largest surveys to date, SEGUE, compiled $\sim50000$ spectra from stars in the solar neighborhood \citep{Yanny2009}.
\citeauthor{Bovy2012} (2012a, hereafter B12) studied a large sample of these and divided the stars into mono-abundance populations (hereafter, MAPs) each of which span a small range in both [$\alpha$/Fe] and [Fe/H].  B12 studied each MAP individually to determine how the structural and kinematic disk properties evolved as a function of element abundance.  B12 made maximum likelihood fits of the disk scale heights and lengths for each MAP and found that each MAP's structure can be described using a single scale length that varies systematically with both [$\alpha$/Fe] and [Fe/H].  The $\alpha$-enhanced MAPs had taller scale heights and shorter scale lengths than MAPs with solar abundance patterns.  This difference implies that the oldest stars formed with shorter scale lengths when the disk was young.  Additionally, the spatial structure of the Galactic disk can be well described by choosing MAPs of approximately solar [$\alpha$/Fe] and showing how their scale lengths depend on [Fe/H]:  more metal-rich MAPs have shorter scale lengths than [Fe/H]-poor ones, which reiterates the presence of a significant radial [Fe/H] gradient in the thin, young disk stellar population.  Finally, the surface density contributions of the scale heights from these populations smoothly decreases from thin to thick MAPs, which lead to the conclusion that the Milky Way is not composed of distinct thin and thick disk populations (\citeauthor{Bovy2012a}, 2012b), a result predicted in theoretical work in which thick disks are the result of radial migration \citep{Schonrich2009,Loebman2011}.  The continuous distribution seems to imply that the thick disk populations did not originate from one significant calamitous event.  

These exquisite data present a new opportunity for comparing models of thick disk formation to observations.  There is a long history of studying the effects of satellite interactions on disk thickening \citep[e.g.][and references therein]{Kazantzidis2008, Kazantzidis2009,Villalobos2010}.  Most such studies have used pure collisionless dynamic simulations and have not included star formation or chemical enrichment modeling, so have focused their comparisons on kinematics and structure, but not chemical enrichment or co-eval stellar populations.  Recently, \citet{Minchev2012} combined a chemical evolution model with sticky particle hydrodynamic simulations and found good agreement between the simulations and observations.  Simulations that have tried to grow disks in a cosmological context have long found a messy beginning for the disks characterized by high early velocity dispersions with thick scale heights \citep{Brook2004,Brook2005} that only settle into thin configurations after evolution \citep{Brook2006}.  Such cosmological hydrodynamic simulations now include star formation and self-consistent chemical evolution that allow more advanced comparisons with observations.  Indeed, \citeauthor{Brook2012} (2012b) made preliminary comparisons of a simulation similar to the one analyzed here and found good general agreement with observations.  \citet{Bird2013} examined the evolution of the disk in the high resolution \emph{Eris} simulation and found that it formed from the inside out and from an initial thick stellar distrubtion to something thinner.

In this paper, we compare a state-of-the-art disk galaxy formation simulation that resembles the Milky Way in its gross properties to the intricate chemo-dynamical patterns observed in the Galactic disk.  If our simulations are realistic, we can explore the validity of the assumptions underlying the interpretation of observations like the assumption that mono-abundance populations (MAPs) are comprised of roughly ``co-eval'' stars.

The simulated galaxy we study here is drawn from the MaGICC project, which simulates galaxies constrained to match the stellar mass--halo mass relationship defined in \citet{Moster2013}.  The MaGICC simulations use supernova feedback and early (pre-supernova) stellar feedback to limit star formation to yield the present-day stellar mass prescribed by the abundance matching technique.  The consequence of the early stellar feedback is to delay star formation in a Milky Way mass galaxy from the typical burst seen in simulations at $z=4$ to a much flatter star formation history that peaks after two gas rich minor mergers that happen at $z=1$ \citep{Stinson2013}.  This history corresponds to a stellar mass to halo mass ratio that evolves similarly to what is found using abundance matching for high redshift luminosity functions \citep{Moster2013,Behroozi2012}.  Star formation is reduced because early stellar feedback maintains gas at 10,000 K and prevents gas from reaching the central 2 kpc of the galaxy \citep{Stinson2013} and supernova feedback ejects the low angular momentum material that would normally create a massive bulge (\citeauthor{Brook2012a}, 2012a).  The resulting galaxy has a flat rotation curve and an exponential surface brightness profile with a scale length of 4 kpc.  

This early stellar feedback has a number of other impacts on galaxy formation.  It drives metal rich outflows that create gaseous halos that match observations of OVI in the circum-galactic medium of star forming galaxies \citep{Stinson2012a}.  It can also expand the inner dark matter density profile of dark matter in galaxies up to nearly L$^\star$, producing dark matter density ``cores" \citep{Maccio2012}.   \citeauthor{Brook2012} (2012b) compared four simulated dwarf galaxies with observations and found that all simulations resulted in good matches to the observed scale lengths, luminosities, and gas fractions.  

In what follows, we present in \S \ref{sec:quantstruct} a quantitative study of the disk structure of the simulated L$_\star$ galaxy, g1536 that was originally described in \citet{Stinson2013}, but is briefly reviewed in \S \ref{sec:sims} along with the physics used in the simulation.  In \S \ref{sec:results}, we show how the disk structure of the simulations compares with observations.  In particular, we compare the correlations between scale heights and lengths and chemical abundances, [$\alpha$/Fe] \& [Fe/H], and the mass weighted disk scale height distribution.

\section{Methods}
\label{sec:sims}
\subsection{Simulations}
We use g1536, a cosmological zoom simulation drawn from the McMaster Unbiased Galaxy Simulations (MUGS).  The simulation starts at $z=99$ from a cosmologically motivated matter distribution based on WMAP3 \citep{Spergel2007}; see \citet{Stinson2010} for a complete description of the creation of the initial conditions.  It includes metallicity-dependent gas cooling, star formation, and a detailed chemical enrichment model that allow us to make a comprehensive comparison with observations of the Milky Way.  

g1536 has a virial mass of $7\times10^{11}$ M$_\odot$, a spin parameter of 0.017, and a last major merger at $z=2.9$, corresponding to a stellar age of $\sim10$ Gyr.  The mass of the Milky Way dark matter halo is uncertain, with values from $7\times10^{11}$ M$_\odot$ to $1.4\times10^{12}$ M$_\odot$ \citep{Klypin2002, Xue2008, Bovy2012d} having been determined using a variety of means.  g1536 is on the low end of this mass range, and thus may not make a perfect comparison to the Milky Way.  However, a comparison may give an idea of which physical processes are important for shaping the structure of a galaxy disk.  The merger history of g1536 is relatively quiet, which may also be the case for the Milky Way \citep{Hammer2007}.  The mergers that do accrete onto the galaxy follow prograde orbits.  Thus, g1536 maintains a disk on a relatively constant spin axis orientation throughout its evolution.  

The simulation uses the smoothed particle hydrodynamics (SPH) code \textsc{gasoline} \citep{Wadsley2004}.  Details of the physics used in the MaGICC project are detailed in 
\citet{Stinson2013}.  Briefly, stars are formed from gas cooler than $T_{max}$ = $1.5\times10^4$ K,
and denser than 9.6 cm$^{-3}$ according to the \citet{Kennicutt1998} Schmidt Law
as described in \citet{Stinson2006} with a star formation efficiency parameter, $c_\star$=0.1.  The star formation density threshold is then set to the maximum density at which gravitational instabilities can be resolved, $\frac{32 M_{gas}}{\epsilon^3}$($n_{th} > 9.3$ cm$^{-3}$), where $M_{gas}=2.2\times10^5$ M$_\odot$ and $\epsilon$ is the gravitational softening (310 pc).

The star particles are $5\times10^{4}$ M$_\odot$, massive
enough to represent an entire stellar population consisting of stars with masses given 
by the \cite{Chabrier2003} initial mass function.  20\% of these have masses greater than
8 M$_\odot$ and explode as Type II supernovae from 3.5 until 35 Myr after the star forms, based on
the Padova stellar lifetimes \citep{Alongi1993, Bressan1993}.  Each supernova inputs 
$E_{SN}=10^{51}$ ergs of purely thermal energy into the surrounding gas.  This energy
would be radiated away before it had any dynamical impact because of the high density
of the star forming gas \citep{Katz1992}.  Thus, the supernova feedback relies on temporarily disabling cooling
based on the subgrid approximation of a blastwave as described in \cite{Stinson2006}.

The stars also chemically enrich the ISM during their evolution through the explosions of SNII and SNIa.  SNII chemical enrichment is based on linear fits as a function of star mass for the \citet{Woosley1995} model of solar metallicity SNII explosions of stars more massive than 8 M$_\odot$.  These models are only based on a single (solar) metallicity.  Linear fits do not suffer from the unrealistic oxygen enrichment that some of the power law fits do \citep{Gibson2002}.  The oxygen fit used is
\begin{equation}
 Y_{Ox}=0.21 M_\star-2 \,\mathrm{M}_\odot
\end{equation}
and the iron fit is
\begin{equation}
 Y_{Fe}=0.003 M_\star 
\end{equation}
SNIa commence enriching the ISM after SNII stop exploding, 40 Myr after the formation of the stellar population, and proceed with nearly constant enrichment for the lifetime of 0.8 M$_\odot$ stars, 12 Gyr.  The SNIa enrichment follows the \citet{Nomoto1984} W7 model from \citet{Thielemann1986}, where each SNIa produces 0.74 M$_\odot$ of iron and 0.143 M$_\odot$ oxygen.

The supernovae feedback does not start until 3.5 Myr after the
first massive star forms.  However, nearby molecular clouds show evidence of being blown apart
\emph{before} any SNII have exploded.  \citet{Pellegrini2007} emphasized the energy input from stellar winds and UV radiation pressure in M17  prior to any SNII explosions and \citet{Lopez2011} found similar energy input into 30 Doradus.  
Thus, in the
time period before supernovae start exploding, we distribute 10\% of the luminosity produced
in the stellar population (equivalent to the UV luminosity) to the surrounding gas without disabling the cooling.  
Most of the energy is immediately radiated away, but \citet{Stinson2013} shows 
that this early stellar feedback has a significant affect on the star formation history
of a Milky Way mass galaxy and places the halo on the \cite{Moster2013} stellar mass--halo mass 
relationship at $z=0$.  

\subsection{Stellar sub-populations}
Throughout the paper, we divide star particles into sub-populations using either their abundance or their age.  We refer to these sub-populations as either a MAP (mono-abundance population) or a co-eval population, respectively.  These are defined as follows:
\begin{itemize}
 \item \emph{Mono-abundance population (MAP)}: Each population is taken from a section of the [O/Fe]-[Fe/H] plane, with $\Delta$[O/Fe]=0.05 and $\Delta$[Fe/H]=0.1.  
 \item \emph{Co-eval}:  Stars all formed at the same time.  The simulation is divided into 50 co-eval populations with equal numbers of star particles. Each population contains 11410 stars and has a characteristic age-spread of $\sim250$ Myr.
\end{itemize}

\subsection{Fitting the structure of sub-population disks}
For comparison with observations, disk structural parameters (scale length and height) are fit throughout this paper using a maximum likelihood method.  Specifically, we fit the number density of simulated star particles following B12, who fit a number density of G dwarf stars.  The radial density profile is modeled with an exponential function while the vertical density distribution uses a $\sech^2(z)$ function to capture the the flattening of the vertical density distribution near the disk midplane.  The likelihood for finding a simulated particle at radius $R_{xy,i}$ and height $|z_i|$ is

\begin{equation}
\begin{array}{llll}
 \ln(\L) = \sum_{i}^{N} -\frac{R_{xy,i}}{r_{exp}} + \ln(\sech^2(\frac{|z_i|}{h_z})) \\ 
- \ln(4\pi r_{exp} h_z (-e^{\frac{-r_{max}}{r_{exp}}}(r_{exp}+r_{max})+ \\
e^{\frac{-r_{min}}{r_{exp}}}(r_{exp}+r_{min})) ) \\
 \left(\tanh(\frac{z_{max}}{h_z})-\tanh(\frac{z_{min}}{h_z})\right),
\end{array}
\end{equation}
where $h_z$ and $R_{exp}$ are the disk structural parameters to be fit in the annulus from $r_{min}=6$ kpc to $r_{max}=10$ kpc in radius and $z_{min}=0$ to $z_{max}=3$ kpc in height.  This likelihood function is turned into a minimization problem by taking its negative.  Initial best fit parameters are found using Powell's method for minimization \citep{Powell1964,Press2007}.  These parameters are used as initial conditions for an ensemble MCMC sampler \citep{Foreman-Mackey2012}.   The MCMC program samples the posterior distribution. The median of the distribution is used as the best fit value because it provides better fits than the results from Powell's method.  The quoted errors are the interval that contains 68\% of the posterior distribution for the given parameter.

\begin{figure*}
 \resizebox{18cm}{!}{\includegraphics{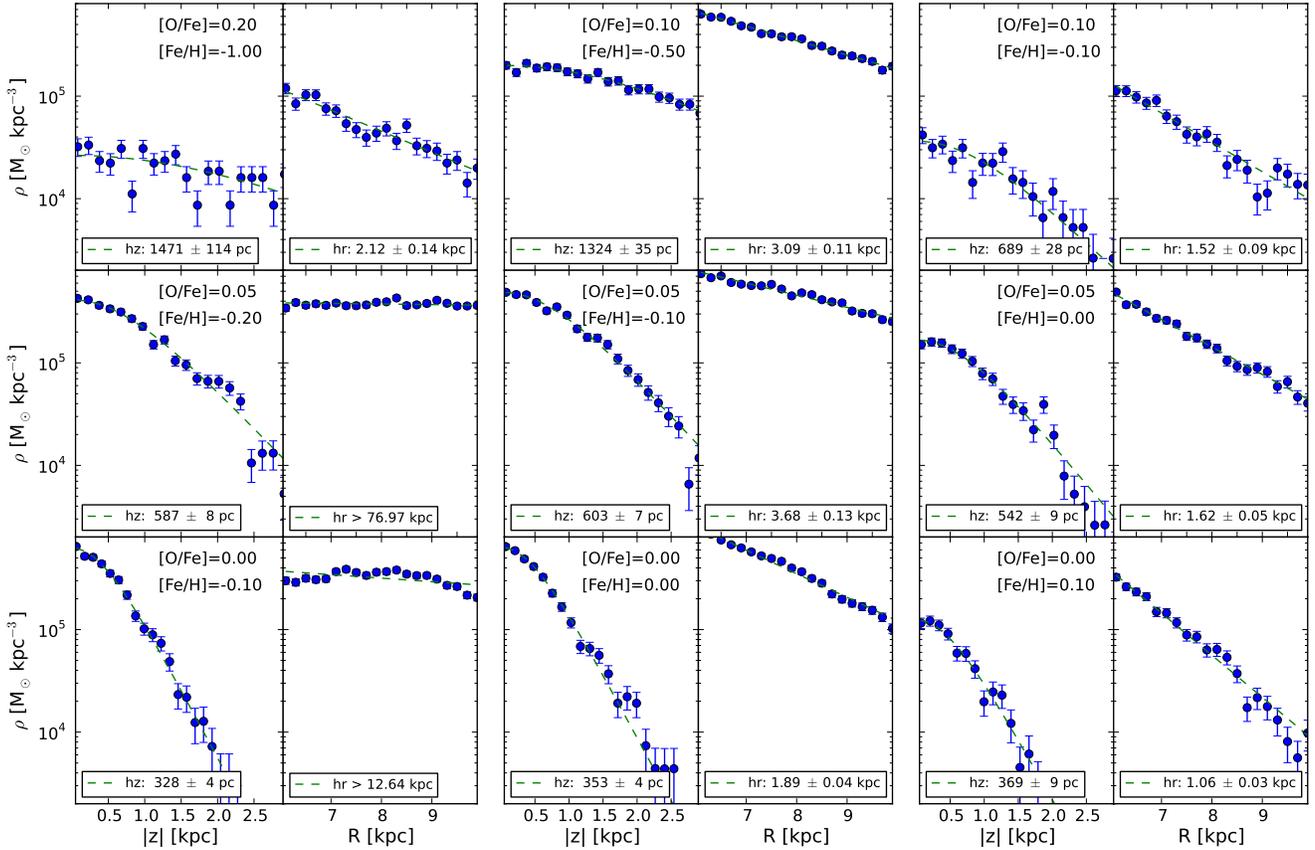}}
 \caption[Metal Bin Fits]{The density profiles for a selection of simulated mono-abundance populations.  This selection is composed of three columns each 2 plots across.  The left plot of the pair is the vertical fit that uses the $sech^2(z)$ function while the right column shows the corresponding exponential radial scale length fit.  The dashed green lines represent the best fit.  The vertical error bars on the points are the Poisson errors.  The top row includes fits from two different [O/Fe] values.  The first pair is for [O/Fe]=0.2-0.25, which shows typical profiles of the $\alpha$-enhanced stellar population.  The second two pairs have [O/Fe]=0.1-0.15 and show some variation in scale length, but to a lesser degree than the bottom two rows.  The plots in the bottom two rows contain a selection of fits with constant [O/Fe] and increasing [Fe/H] from left to right.  The bottom row is [O/Fe]= 0-0.05, the middle row is [O/Fe]=0.05-0.1.  These [O/Fe] values are selected because of the large variation in scale length as a function of [Fe/H] as shown in Figure \ref{fig:blockofefeh}.  In several cases, the stellar density is nearly flat as a function of radius.  In these cases, the quoted scale lengths are taken from the bottom 1\% of the posterior distribution so represent an extreme lower limit of the scale length.  }
 \label{fig:metfits}
\end{figure*}

The profiles for a selection of example MAPs are presented in Figure \ref{fig:metfits}.  It shows the great variation of profiles apparent in the various MAPs, differences that will be described in \S \ref{sec:structure}.  The models provide a good description of the data.

\section{Qualitative Structural Evolution}
\label{sec:quantstruct}
\begin{figure*}
  \subfigure[where stars formed]{
    \includegraphics[width=0.48\textwidth]{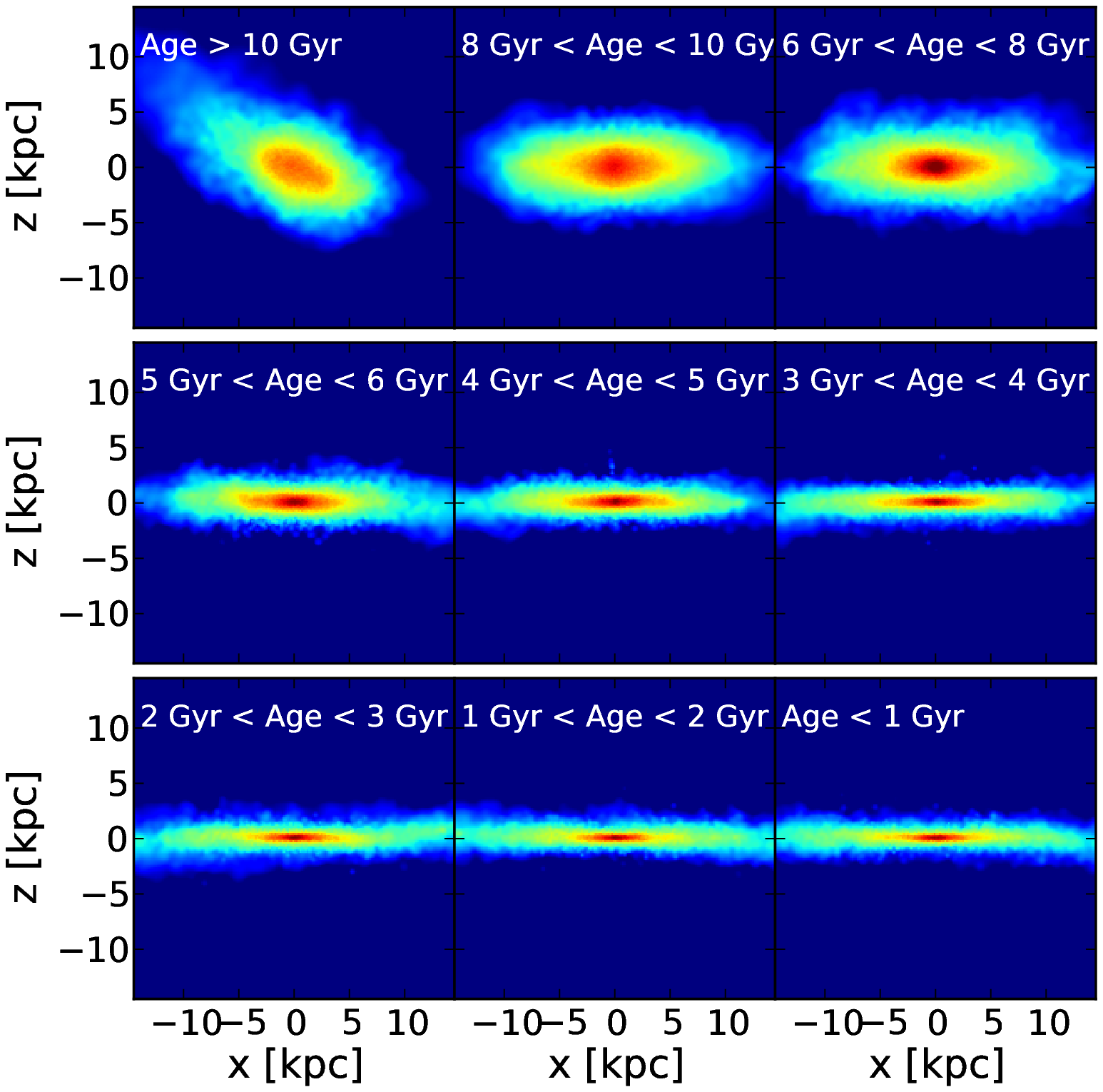}
  } 
  \subfigure[where stars are at $z=0$ (co-eval populations)]{
    \includegraphics[width=0.48\textwidth]{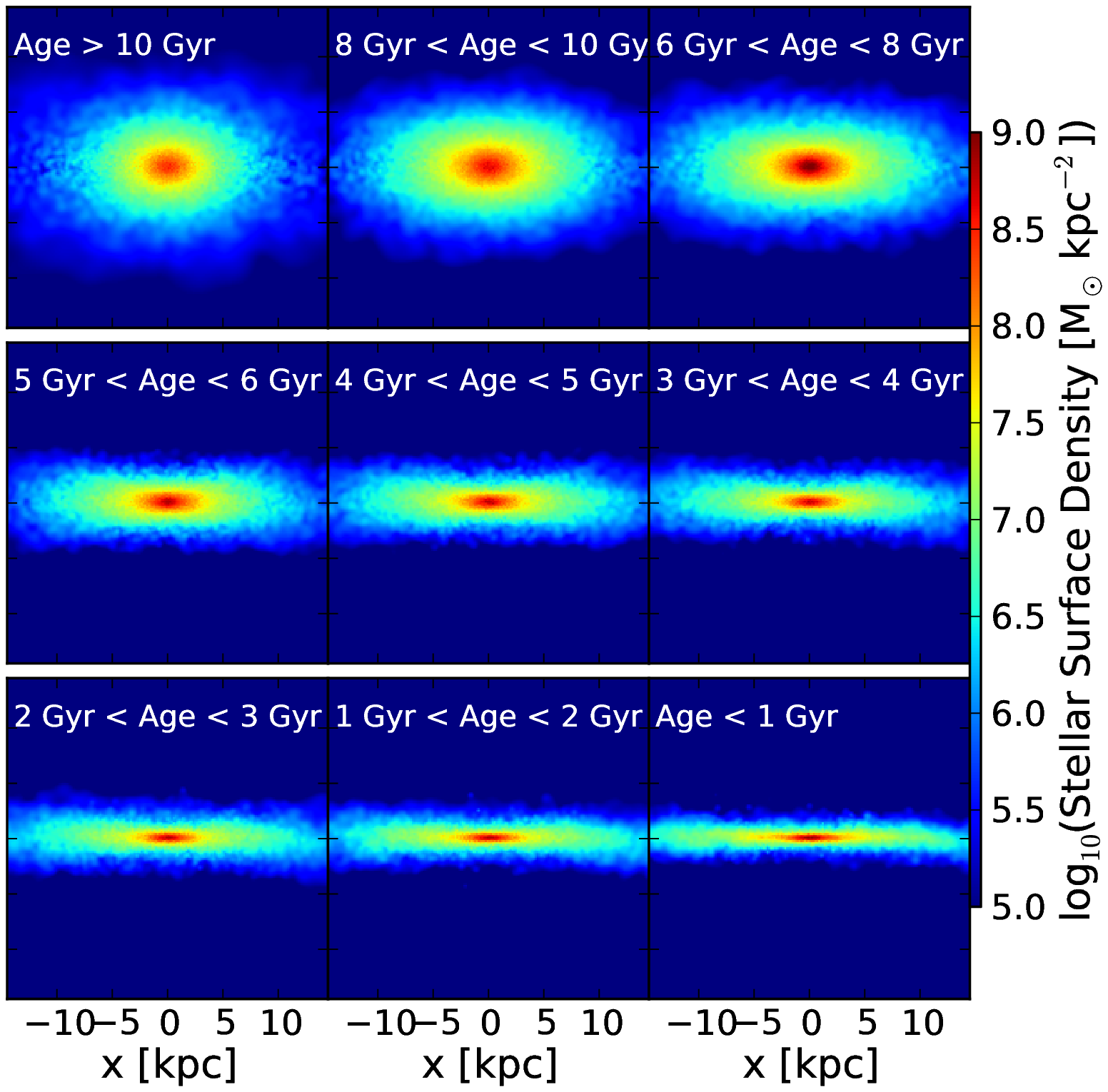}
  } 
 \caption[Edge-on stellar density]{ Projections of the stellar density for stars grouped into nine different age bins.  The left panel shows how the location of the stars when they formed.  The right panel shows the present-day ($z=0$) distribution of the various MAPs.  In both cases, there is a trend from round (large scale height, short scale length) to flattened (small scale height, long scale length) distributions, when going from old to young sub-populations. }
\label{fig:edgeonev} 
\end{figure*}
To give a brief introduction to the analysis that follows, we begin with images of the stellar populations that comparise our simulated galaxy.
Figure \ref{fig:edgeonev} shows the evolution of various co-eval populations in the simulation, panel (a) shows where the stars formed, while panel (b) shows where those same stars are at $z=0$.  These images give a qualitative sense that the oldest stars formed with a short scale length and tall scale height, while the youngest stars have a long scale length disk with a short scale height.  A comparison with the right panels shows that the co-eval populations have not evolved significantly in their structure other than some slight thickening.  Thus, it is possible to look at the structural properties of present day stellar populations and make inferences about where those stars formed.

\begin{figure}
 \includegraphics[width=0.48\textwidth]{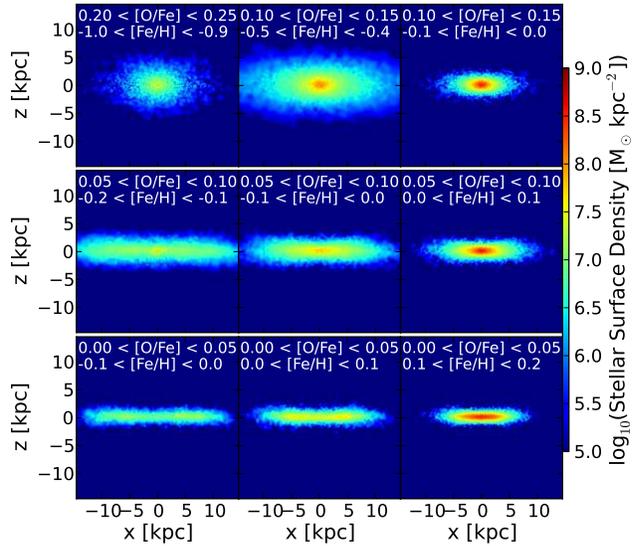}
 \caption[MAP structures]{Edge-on stellar surface densities for a selection of mono-abundance stellar populations (MAPs).}
 \label{fig:mappics}
\end{figure}
In \S \ref{sec:results}, we examine the structural parameters of mono-abundance populations (MAPs).  Figure \ref{fig:mappics} shows the $z=0$ stellar structure for a sample of MAPs.  The selection includes representatives from the three main MAP families we find in our quantitative analysis in \S \ref{sec:results}: old-thick disk, intermediate, and young-thin.  The upper-left panel represents the old-thick population that has a tall scale height, short length and enhanced [O/Fe].  The right two panels in the upper row represent the intermediate population that arises from a significant transition in the evolution of the stellar disk.  In this population, the stars with the tallest scale heights also have the longest scale lengths.  The bottom two rows represent the young-thin population.  All of these populations have as thin a disk as can be resolved in the simulation.  The scale lengths, however, vary greatly as a function of [Fe/H], creating a significant [Fe/H] gradient that is apparent in the youngest, solar [O/Fe] abundance populations.

\section{Results}
\label{sec:results}
With the procedure in place for fitting the scale height and length of sub-samples of stars in the simulations at $z=0$, we can now compare the mono-abundance populations (MAPs) between the observed Galactic disk and the simulations.  We further study how the structure of the MAPs compares to co-eval populations in \S \ref{sec:agestruct}.
\subsection{Simulated MAP Structure}
\label{sec:structure}
\begin{figure*}
  \subfigure[Scale height]{
    \includegraphics[width=0.48\textwidth]{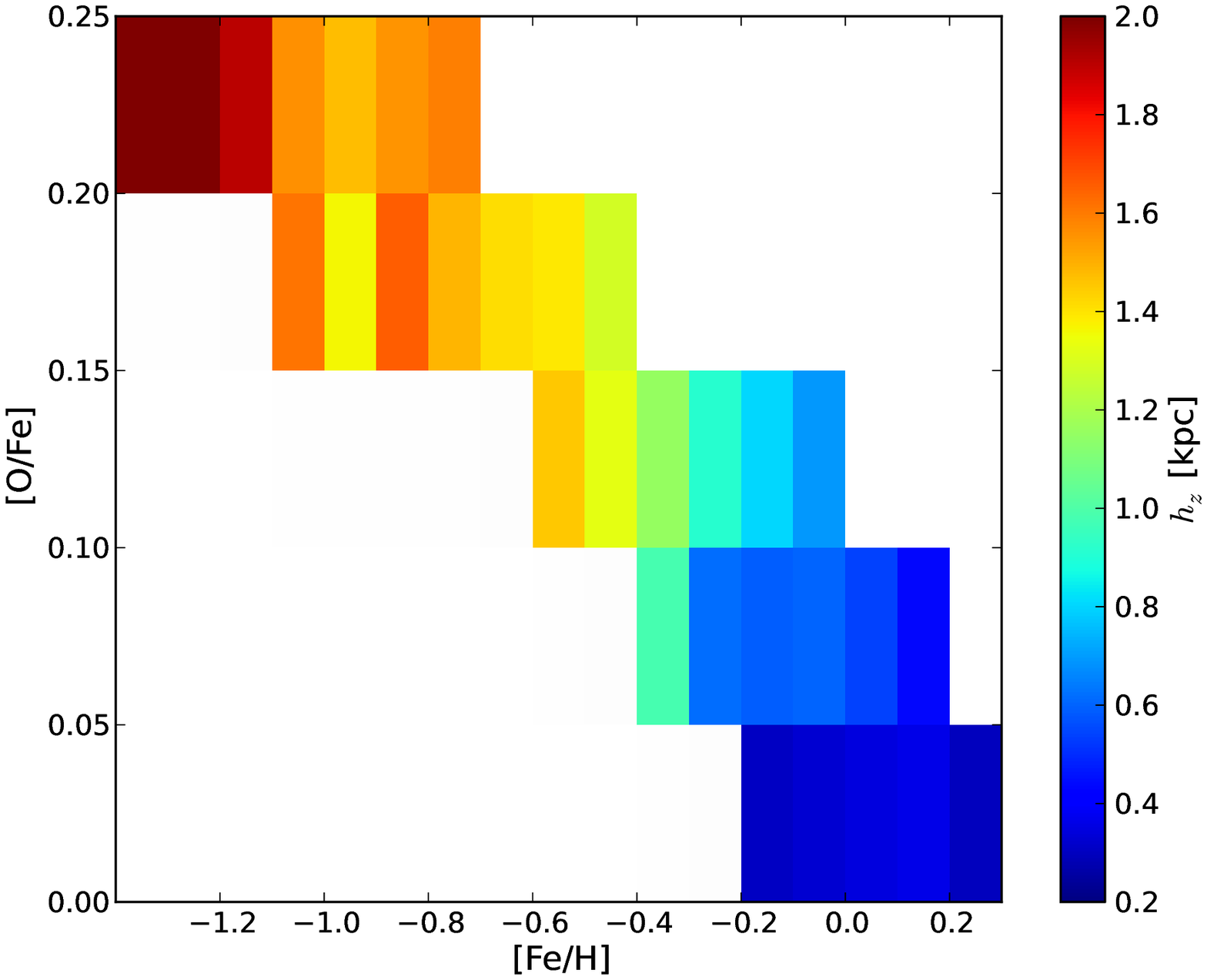}
  } 
  \subfigure[Scale length]{
    \includegraphics[width=0.48\textwidth]{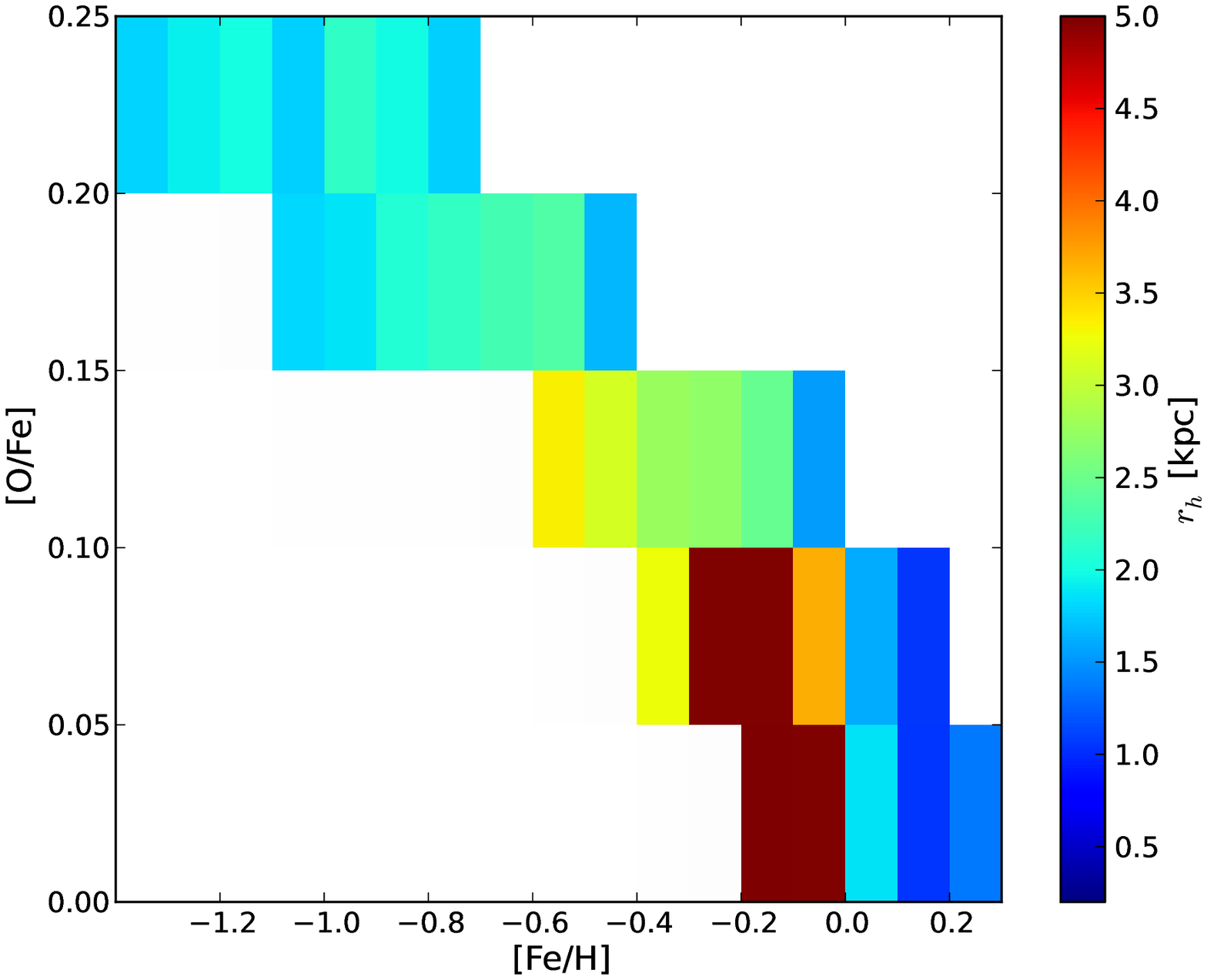}
  } 
 \caption[Scale Height and Lengths]{ Dependence of the scale height (left) and length (right) as a function of [Fe/H] and [O/Fe] for different simulated mono-abundance stellar populations.  Stars are divided into abundance populations with $\Delta$[O/Fe]=0.05 and $\Delta$[Fe/H]=0.1. These are the same fits as Figure \ref{fig:hzrexp} only projected into the abundance plane.}
\label{fig:blockofefeh} 
\end{figure*}
Figure \ref{fig:blockofefeh} shows how the disk scale height (left) and length (right) vary as a function of [Fe/H] and [O/Fe] for different MAPs in the simulation.  There are smooth changes in both the scale height and length as a function of abundance.  The tallest scale heights exist in the [Fe/H]-poorest and most oxygen-enhanced populations.  The scale height shrinks as the [Fe/H]-enrichment increases and the oxygen-enhancement decreases, reaching a minimum of $\sim~200$ pc.  

Regarding the scale lengths in the right panel, the oxygen-enhanced stellar populations all have comparable scale lengths around 2 kpc.  In less oxygen-enhanced populations, the scale length generally grows.  However, the [Fe/H]-enriched stellar populations at solar [O/Fe] have scale lengths that are shorter, $r_{exp}\sim$1 kpc, than the oxygen-enhanced populations.  

The [Fe/H]-enriched populations represent the most recent star formation, occurring near the galactic center.  There is a limited radius inside which the ISM has been sufficiently enriched to produce such stars.  Outside this radius, star formation is also ongoing, but with a scale length much larger than fits on the colorbar scale.  The change in scale length is separated by a variation of only 0.2 dex in [Fe/H].  

The $\alpha$-enhanced populations show much less variation in their scale lengths, staying around 2 kpc.  They show only moderate variation in their scale heights, which gradually decrease with higher [Fe/H] and lower [O/Fe].  These are the stars that make up the thick disk of g1536.  \citet{Liu2012} found that in the Milky Way, stars with $\alpha$-enhanced abundances share many kinematic properties.

The range in [O/Fe] abundances is not as wide in g1536 as in the Milky Way (0-0.25 versus 0-0.5).  This difference may indicate a problem with the exact treatment of chemical enrichment in the simulation. 

\begin{figure}
    \includegraphics[width=0.48\textwidth]{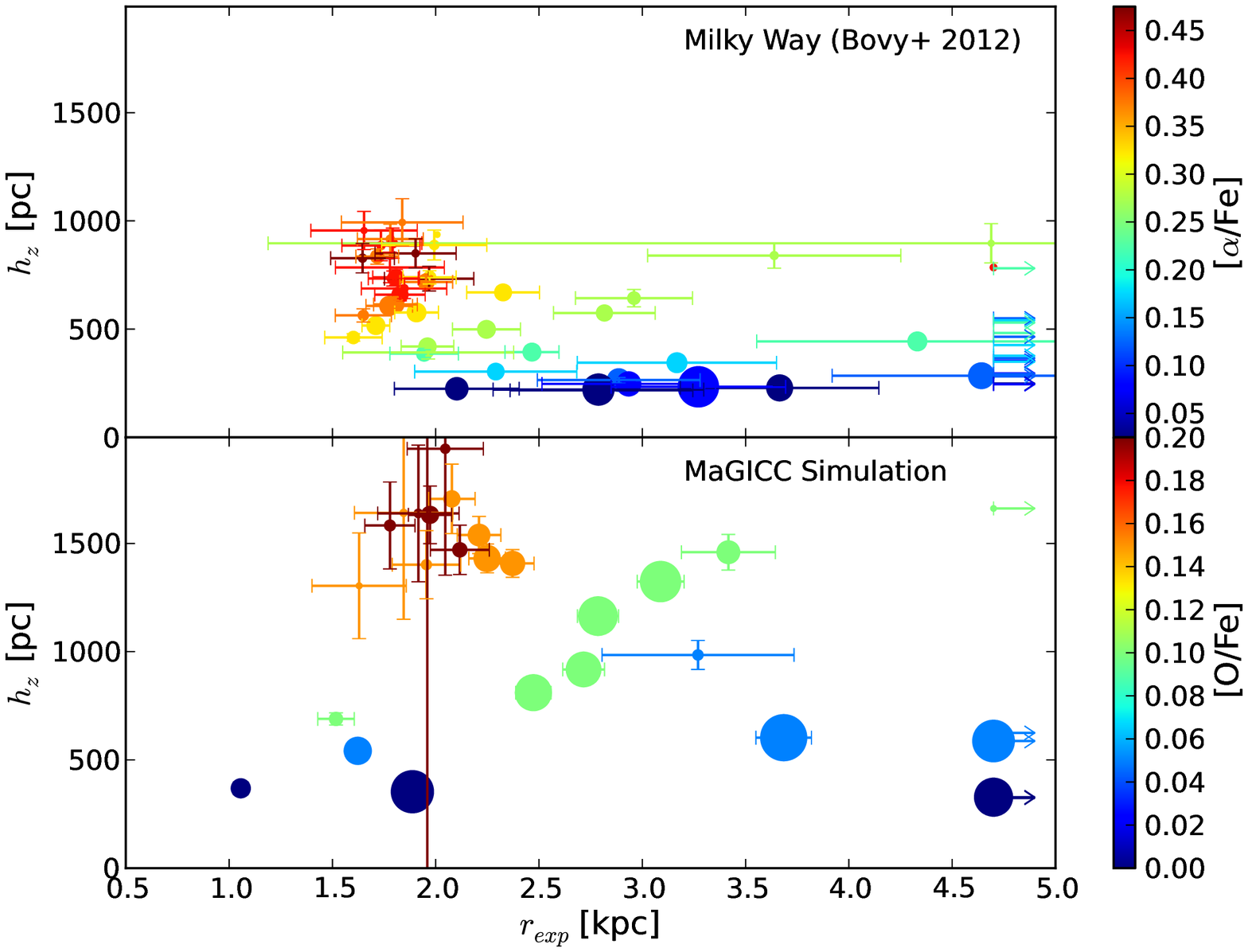}
 \caption[Shape evolution]{ Present-epoch scale height vs scale length for various MAPs in the simulations.  The stars are split into mono-abundance populations, 0.05 wide in [O/Fe] and 0.1 in [Fe/H].  The simulated points are coloured by [O/Fe], the observed points are all grey.  In many cases, the errors are smaller than the symbol size.  The simulated MAPs show some trends. The scale height decreases with [O/Fe], while the scale length remains around a constant value of 2 kpc.  Below a scale height of 600 pc, the scale lengths are longer. }
\label{fig:mbhzrexp} 
\end{figure}

Figure \ref{fig:mbhzrexp} details the relation between the scale height and the scale length for the simulated MAPs (bottom panel) and compares it to the Galactic observations from B12 (top panel).  The distribution of simulated MAPs shows some structure.  There is a cluster of $\alpha$-enhanced points with short scale lengths and large scale heights, comparable to the cluster of $\alpha$-enhanced points in the top panel of Figure \ref{fig:mbhzrexp} (or Figure 5 of B12). 

Figure \ref{fig:mbhzrexp} illustrates an important qualitative difference between the simulation and the Galactic disk: the simulated disk is $\sim$50\% thicker than the Milky Way.  The extra thickness is correlated with the $\sim50$\% greater velocity dispersion in the simulation ($\sigma_z\sim$60 km s$^{-1}$ in solar neighborhood for the oldest, thickest population, $\sigma_z\sim$30 km s$^{-1}$ for a 4 Gyr old population).  The variation in each may be due to either a difference in mass between g1536 and the Milky Way or stellar feedback that is too strong.  As mentioned before, g1536 is at the low end of the Milky Way total mass estimates and has less stellar mass than the Milky Way disk.  Less mass would make the disk potential shallower and thus the disk thicker at a given velocity dispersion.  

Switching focus to the MAPs with low (solar) [O/Fe], Figures \ref{fig:blockofefeh} \& \ref{fig:mbhzrexp} show that they all have scale heights $<500$ pc and a wide range of scale lengths from 1 kpc to large values.  Figure \ref{fig:metfits} shows that the low [Fe/H] bins have the longest scale lengths, such that the fraction of stars with such low [Fe/H] values increases further out in the disk.  Figure \ref{fig:metfits} shows that the stellar density profiles are nearly flat as a function of radius for these populations. As [Fe/H] increases, the scale length becomes dramatically shorter.  Such a scale trend with [Fe/H] is also clear in Figure 5 of B12 and points to a metallicity gradient present in recently formed stars.  

Our simulation predicts the existence of a compact metal-rich stellar population with short scale height and length.  It is plausible that this population did not show up in the analysis of B12 because the SDSS G-dwarf data sets do not include stars within 300 pc of the disk midplane or sightlines towards the Galactic centre.

There are also intermediate (0.1-0.15) [O/Fe] stellar populations that show a direct correlation between scale height and length.  Figure \ref{fig:blockofefeh} shows that this is again related to the metallicity gradient of the disk.  The most enriched MAP ([Fe/H] $\sim0$) in this [O/Fe] range has the shortest scale length and height of these populations.  At progressively lower [Fe/H] abundances, the scale height and length both become longer.  This correlation is also apparent in Figure 5 from B12, although only in the observed MAPs where fits are poor.  We will examine this intermediate population further in the stellar populations summary, \S\ref{sec:popsum}.

\subsection{Radial abundance gradients}
\begin{figure}
 \resizebox{9cm}{!}{\includegraphics{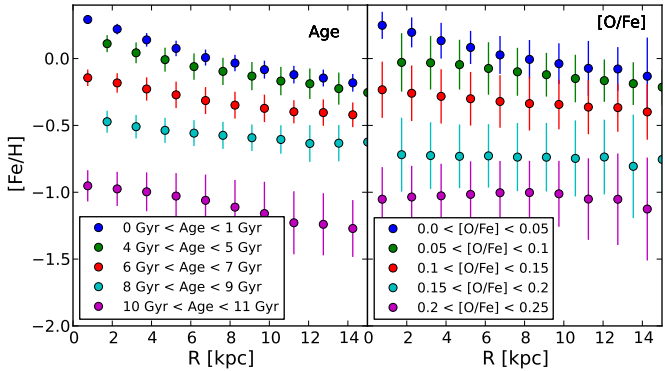}}
 \caption[Metallicity Gradient]{The metallicity gradient of stellar populations when split either by age (left panel) or by $\alpha$-enrichment [O/Fe].  Each plot shows the mean [Fe/H] in 1.5 kpc radial bins. The radii of alternating populations are offset by 750 pc so that the error bars can be seen. The vertical error bars represent the spread (variance) of the [Fe/H] distribution in each bin.}
 \label{fig:fehgradient}
\end{figure}

To make the [Fe/H] radial gradient clearer and identify the best way to search for its signature in observations, Figure \ref{fig:fehgradient} shows the [Fe/H] gradient in populations divided using either age or [O/Fe].
As expected from Figures \ref{fig:blockofefeh} and \ref{fig:mbhzrexp}, there is a strong radial metallicity gradient in populations that are relatively young, or have solar [O/Fe] values (blue symbols), with $\Delta$[Fe/H]/$\Delta$R$\sim0.04$ dex/kpc.  In contrast, the old or [O/Fe]-enhanced stars (purple symbols) barely have any radial [Fe/H] gradient.  Thus, the [O/Fe]-separated [Fe/H] gradient is a good proxy for the gradient that appears in the youngest populations.  

However, the lack of a significant gradient in the old stellar populations that comprise the thick disk means that the [Fe/H] gradient becomes less clear when divided by scale height rather than when stars are divided by age or [O/Fe].  The significant scatter at large scale heights makes it difficult to use [Fe/H] gradients to identify what mechanism creates thick disks. 

\subsection{Comparing abundances and ages}
\begin{figure*}
  \subfigure[Mean Age]{
    \includegraphics[width=0.48\textwidth]{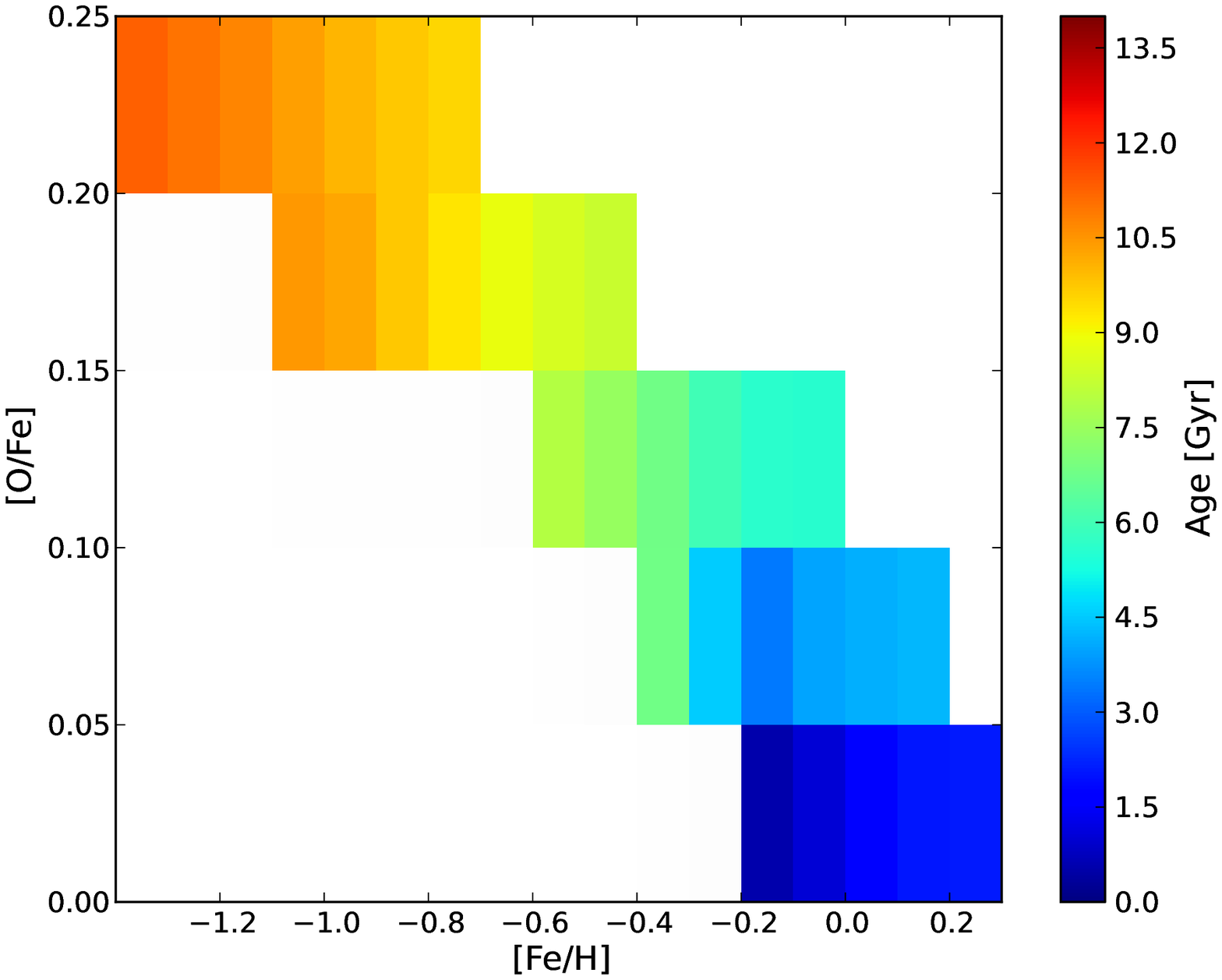}
  } 
  \subfigure[Age Dispersion]{
    \includegraphics[width=0.48\textwidth]{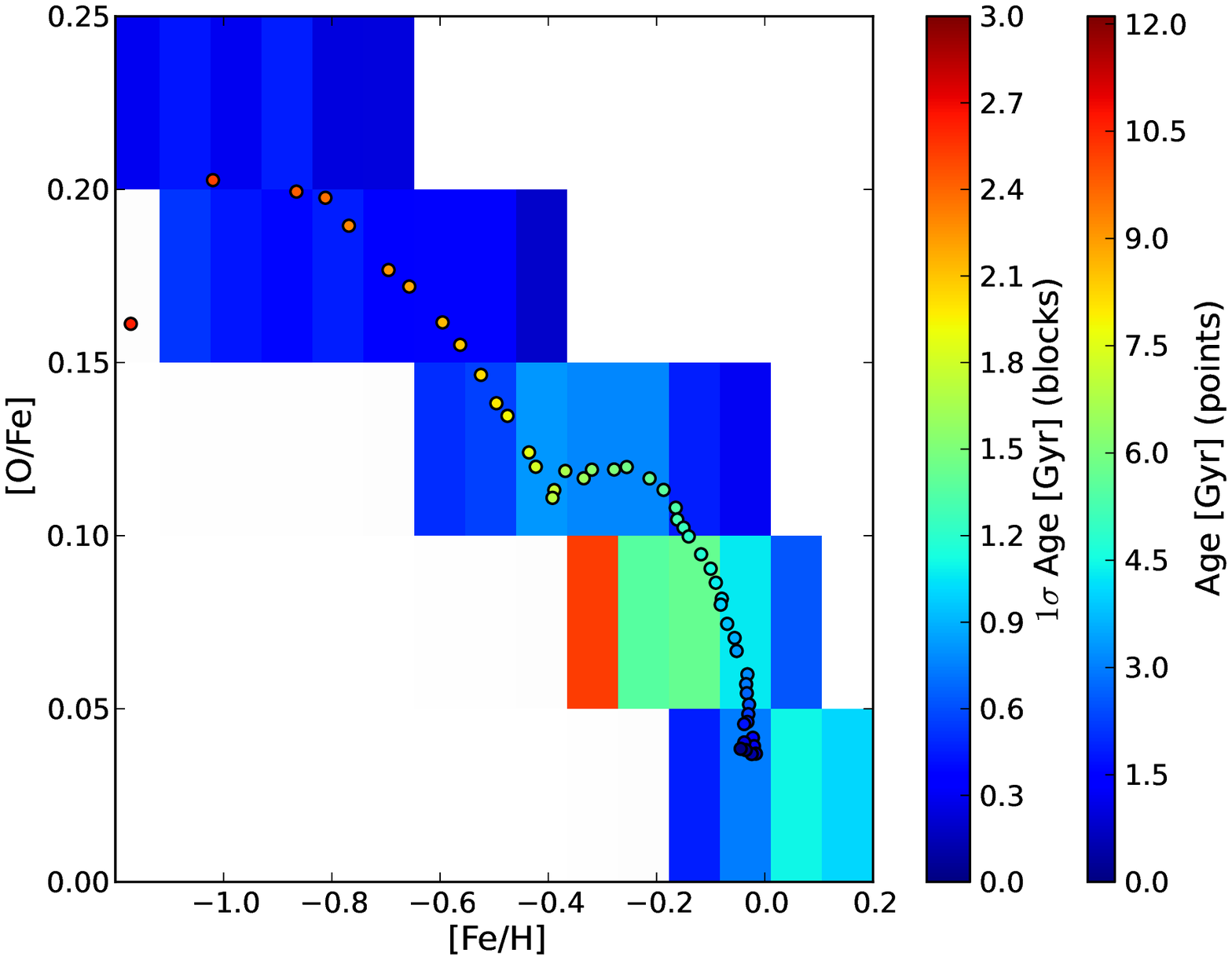}
  } 
 \caption[Age Dispersion]{(a) The mean age of the simulated mono-abundance populations; (b) the age dispersion (central 68\% of the age distribution) for each mono-abundance population.  The dots in the right panel are the mean [O/Fe] and [Fe/H] values for stars binned by their age.  The age dispersion is less than 1 Gyr for nearly every MAP except [O/Fe]=0.05-0.1, [Fe/H]=-0.3 to -0.2.  This abundance corresponds to the time when the abundance evolution moves to a trajectory at higher [O/Fe] due to a sudden increase in star formation. }
\label{fig:blockages} 
\end{figure*}

In the simulation, we can examine the age-distribution of the stars within each MAP.  The left panel of Figure \ref{fig:blockages} shows the mean age for all the MAPs.  It shows that the [O/Fe]=0.1-0.15  MAPs formed between 5 and 7.5 Gyr ago.  In this range of [O/Fe] values, the stars that formed the longest ago have the longest scale length and height.  This is an indication that the disk did not strictly form from the inside out.  This intermediate [O/Fe] population seems to indicate that there is a transition epoch as the disk shifts from a short, thick distribution to a long, thin distribution.

The right panel of Figure \ref{fig:blockages} shows, however, that most stars within each MAP did form at similar times, with a typical age spread of less than 1 Gyr.  This implies that [O/Fe] {\it and} [Fe/H] jointly are a good proxy for co-eval populations across much of the [O/Fe]-[Fe/H] abundance plane.

The MAP with 0.05$<$[O/Fe]$<$0.1 and -0.3$<$[Fe/H]$<$-0.2 has an exceptionally large dispersion, containing contributions from stars with a range of 6 Gyr.  We overplotted the mean [O/Fe] and [Fe/H] values for stars split into age bins.  These points show that the normal enrichment pattern makes an excursion towards higher $\alpha$-enrichment about 7 Gyr from the end of the simulation.  A look at the star formation history of this galaxy in Figure 9 of \citet{Stinson2013} shows that this is the time at which the galaxy undergoes an increase in star formation.  An examination of the evolution of the galaxy\footnote{http://www.mpia.de/~stinson/magicc/movies/c.1td.05rp.1/far.mp4} shows that there are two prograde minor mergers in rapid succession during this epoch.  This increase in star formation causes stars to form with higher [O/Fe] and then continue on a parallel abundance trajectory.  No [$\alpha$/Fe] excursion has yet been observed in the Milky Way.

\subsection{Structure of the Co-eval Populations}
\label{sec:agestruct}
We repeat our analysis of the evolution of structural parameters, but using co-eval population rather than MAPs.
\begin{figure}
    \includegraphics[width=0.48\textwidth]{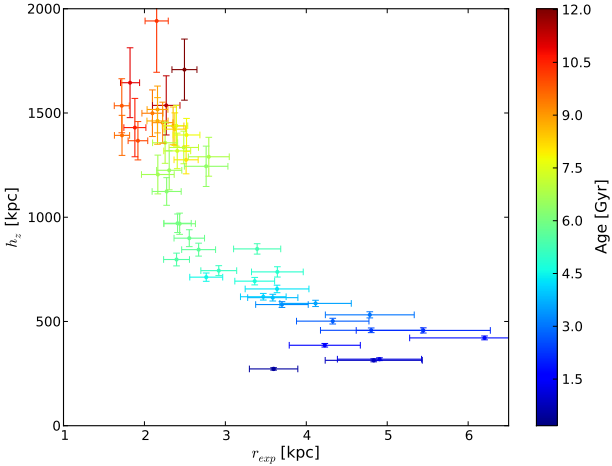}
 \caption[Shape evolution]{ Present-epoch scale height vs scale length for various co-eval populations in the simulations.  The stars are split into 50 equal sized age bins.  There is an obvious trend when using the age binning.  The scale height decreases gradually, while the scale length remains around a constant value of 2 kpc.  When the scale height decreases to 600 pc, the scale length starts to grow.  }
\label{fig:hzrexp} 
\end{figure}
Figure \ref{fig:hzrexp} shows the scale height, $h_z$, as a function of scale length, $r_{exp}$ for 50 different co-eval populations between $r_{min}=6$ kpc and $r_{max}=10$ kpc.  The structural evolution of the co-eval populations is smoother and more gradual than for the MAPs.  The oldest stars are in a thick distribution with a scale length of 2 kpc.  The younger populations have progressively shorter scale heights, and longer scale lengths.  For these young stars, the disk scale length begins to increase while the scale height becomes more constant.  

Compared to the MAPs in Figure \ref{fig:mbhzrexp}, the wide range of scale lengths for the recently formed, solar [O/Fe] MAPs disappear.  Instead, the scale lengths of the youngest population are 4-6 kpc, an average of the wide range in the MAPs.  Figure \ref{fig:fehgradient} shows that the [Fe/H] gradient is still quite strong in the youngest population, even stronger than for the MAPs since the MAPs mix populations of different ages.  What is apparent from a comparison of the structure of the co-eval and mono-abundance populations is that the structure of MAPs is a mixture of disk evolution and enrichment.  The enrichment is not uniform.  

\begin{figure}
 \resizebox{9cm}{!}{\includegraphics{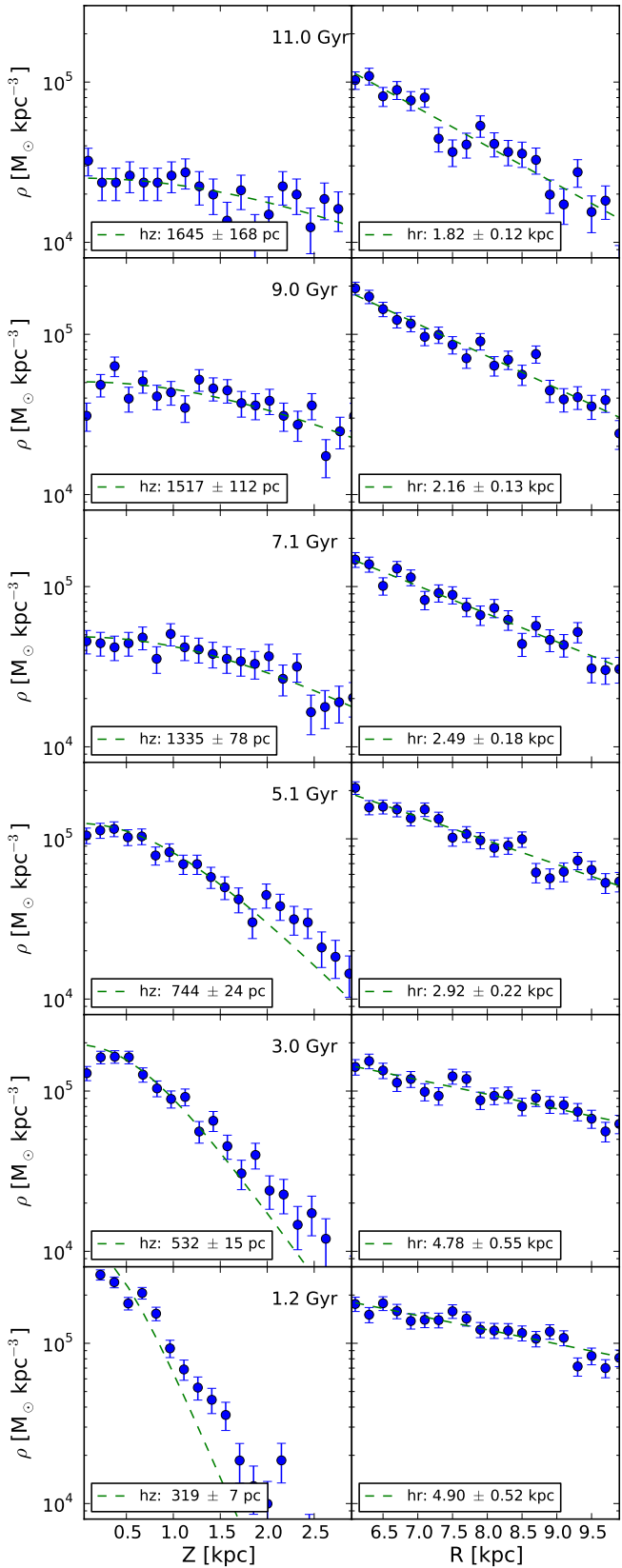}}
 \caption[Age Fits]{Scale height and scale length of the simulated co-eval stellar populations.  The left column is the vertical fit of a $\sech^2(z)$ profile while the right column shows the corresponding radial exponential scale length fit.  The vertical error bars on the points are the Poisson errors.}
 \label{fig:agefits}
\end{figure}
Figure \ref{fig:agefits} shows the fits of the disk scale height and length of stars in an annulus 6 to 10 kpc from the galactic center, comparable to the plots in Figure \ref{fig:metfits}.  These sample fits show the gradual transition from a thick disk with a short scale length to a thin disk with a long scale length.

\subsection{The Mass-Weighted Scale Height Distribution}
\begin{figure}
 \resizebox{9cm}{!}{\includegraphics{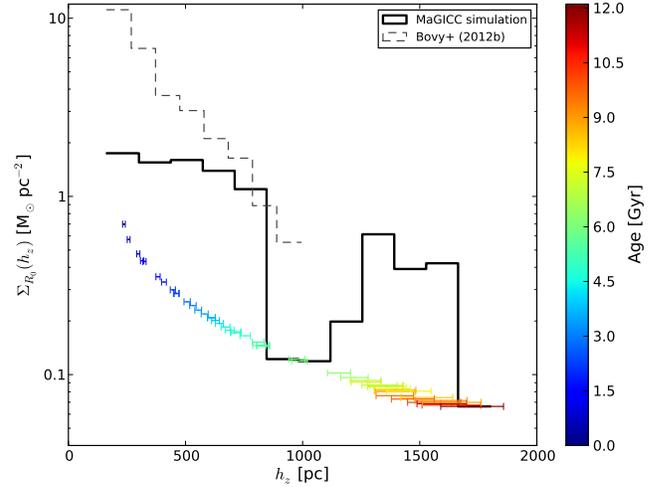}}
 \caption[hz Surface Density Distribution]{Surface density at the solar radius (8 kpc) as a function of scale height for coeval populations.  The surface densities for the points colored by age were calculated using the scale height and length of each co-eval population in the density equation integrated over $|z|$ from 0 kpc to 3 kpc.  The points are then placed according to their scale height.  The solid line represents how the surface densities of the simulated galaxy are distributed across 150 pc bins.  The dashed line represents the same distribution of surface densities for MAPs taken from \citeauthor{Bovy2012a} (2012b).}
 \label{fig:surfdens}
\end{figure}

We can calculate how much different sub-populations contribute to the local surface density at the solar circle.  For each co-eval population, we calculate its integrated surface density using its scale height and length in the density equation :
\begin{equation}
 \rho(R,z) = \rho_0 e^{\frac{-R}{R_{exp}}} \sech^2\left(\frac{|z|}{2 h_z}\right)
\end{equation}
integrated over $|z|$ from 0 to 3 kpc:
\begin{equation}
 \Sigma_{R\odot}(h_z) = 2\rho_0 e^{\frac{-R_\odot}{R_{exp}}} \int_{0\,\mathrm{kpc}}^{3\,\mathrm{kpc}} \sech^2\left(\frac{|z|}{2 h_z}\right) dz
\end{equation}

We then plot this surface density as a function of the characteristic scale heigh of each co-eval population, which roughly maps to metallicity and abundance (Figure \ref{fig:blockages}).  Figure \ref{fig:surfdens} shows that there is a nearly flat distribution of the surface density contributed by populations with large scale heights up to 800 pc.  There is then an absence of surface density contributed by populations with scale heights between 800 and 1200 pc, but then a distinct component with scale heights from 1.2 kpc to 1.5 kpc, which corresponds to the old, thick disk formed in the simulation.  These scale heights are far greater than any so far observed in the Milky Way and may be due to the low mass of g1536 or the overheating from the stellar feedback.  The absence of populations with scale heights around 1 kpc illustrates how the simulation transitions suddenly from it thick disk forming phase to thin disk.  The data from the Milky Way does not show any such sudden transition.

\subsection{Summary of Populations}
\label{sec:popsum}
Based on the comparison of our simulations with observations, we identified three distinct stellar populations.  The simulation shows clear indications of the evolution from one population to the next as the disk evolves and the signature of these transitions are also apparent in the observed MAPs.

\subsubsection{Old thick disk}
The MAPs with [O/Fe]$>$0.15 all have constant scale lengths of $\sim2$ kpc in both the simulated galaxy and the Milky Way.  These MAPs also have the tallest scale heights, though they shrink by a factor of two as the disk evolves.  The MAPs are all older than 8 Gyr and over the 5 Gyr of their evolution, their [Fe/H] increases from -1 to -0.5 in both the simulated galaxy and the Milky Way.  The shrinking scale height is steady as [Fe/H] becomes more enriched in the Milky Way.  While the shrinking scale height is not so clear as a function of [Fe/H] in the simulated galaxy, it is apparent as a function of age in Figure \ref{fig:hzrexp}.

\subsubsection{Intermediate population}
There is a general evolution of scale height as an inverse function of scale length.  There is one series of MAPs that does not follow this trend, but instead scale height directly follows scale length.  These MAPs all have [O/Fe]=0.1-0.15 in the simulations and 0.25 in the observations.  However, this trend does not show up at all in Figure \ref{fig:hzrexp}, which displays scale height as a function of scale length for co-eval populations.  It shows a strictly inversely proportional evolution.  Figure \ref{fig:blockages} shows that the reason these MAPs show the odd trend is because they include the widest range of ages.  These MAPs are filled with stars $\sim$6 Gyr old that formed when the abundance evolved to a more [Fe/H]-enriched track.  This abundance evolution happens due a suddenly increased star formation rate.  These populations are much more massive in the simulation than in the Milky Way.

\subsubsection{Young thin disk}
At the lowest [O/Fe] enhancements, the disks in both g1536 and the Milky Way are their thinnest.  For g1536, these populations have [O/Fe]$<0.1$ and [Fe/H]$>-0.5$, while in the Milky Way, they have [O/Fe]$<0.2$ and [Fe/H]$>-0.7$.  The scale heights of these populations range from 500 down to 200 pc in the Milky Way and from 700 down to 300 pc in g1536 with the lower [O/Fe] enhanced populations have the thinnest disks.  The scale lengths vary greatly for these MAPs from 2 to $>5$ kpc in the Milky Way and from 1 to $>10$ kpc in g1536. In both cases, these scale lengths vary directly as a function of [Fe/H].  All these MAPs are $<3$ Gyr old in the simulation, so the wide variation in [Fe/H] represents the metallicity gradient evident in the young thin disk.

\subsubsection{Differences between simulation and Milky Way}
The simulated galaxy has scale heights that are twice those observed in the Milky Way.  The simulation also shows evidence for a sudden event that makes the disk much thinner over a short period of time. It leaves an absence of stars that have a scale height of 1 kpc in the distribution of scale heights.  The Milky Way plot made from MAPs shows no similar absence, so it appears that the MW more smoothly evolved from thick to thin disk formation.

\section{Conclusions}
We make a detailed comparison between the stellar structure of a disk formed in a cosmological simulation and recent observations of the chemical and kinematic structure of the Galactic disk.  In particular, we compare the systematic variations of disk scale height and lengths between simulations and observations.  At a qualitative level, we find a striking similarity.  

Our disk structural analysis is based on both age and chemical enrichment.  The oldest stellar populations have short scale lengths and tall scale heights.  When the the disk is decomposed according to metal abundance as in observations, the thickest populations that have the short scale length are those that are $\alpha$-enhanced.  It is clear in the simulations that these are the oldest populations.  The young stellar populations that have solar $\alpha$-enrichment have the shortest scale heights.  Since we have divided our populations in [O/Fe] and [Fe/H] simultaneously, we can see that these young populations have a wide range of scale lengths that is strongly dependent on [Fe/H].  This illustrates how much chemical enrichment varies at different locations in the disk.  In particular, the center of the galaxy is the most the metal ([Fe/H]) enriched while the outskirts of the disk follow a lower [Fe/H] path to solar [O/Fe].

The correspondence between the Milky Way and the MaGICC simulation add support to scenarios where thick disks form kinematically hot.  In future work, we will show how the morphology of the MaGICC galaxy evolves, but since it is a relatively low mass galaxy, it does not have large clumps of star formation at high redshift.  Rather, the stellar feedback heats the galactic gas early in its evolution so that it does not settle into any sort of thin disk until $z\sim1$.  The remaining presence of this thickened stellar component is what compares best with the thick disk observed in the Milky Way.  As the galaxy increases in mass the gas settles down and forms a thin disk.

\section*{Acknowledgements}
We thank the anonymous referee for helpful comments that have greatly improved this paper.
The analysis was performed using the pynbody package
(\texttt{http://code.google.com/p/pynbody}), which was written by Andrew Pontzen
in addition to the authors.
We thank Peter Yoachim, Sarah Loebman, Connie Rockosi, and Neil Crighton for useful conversations regarding this paper.
The simulations were performed on the \textsc{theo} cluster of  the
Max-Planck-Institut f\"ur Astronomie at the Rechenzentrum in Garching;
the clusters hosted on \textsc{sharcnet}, part of ComputeCanada.  We greatly appreciate
the contributions of these computing allocations.
J.B. was supported by NASA
through Hubble Fellowship grant HST-HF- 51285.01 from the Space
Telescope Science Institute, which is operated by the Association of
Universities for Research in Astronomy, Incorporated, under NASA
contract NAS5-26555. J.B. was partially supported by SFB 881 funded by
the German Research Foundation DFG and is grateful to the Max-Planck
Institut f\"{u}r Astronomie for its hospitality during part of the
period during which this research was performed.
CBB acknowledges Max- Planck-Institut f\"ur Astronomie for its hospitality and financial support through the  Sonderforschungsbereich SFB 881 ``The Milky Way System''
(subproject A1) of the German Research Foundation (DFG).  AVM and GS also acknowledge support from SFB 881 (subproject A1) of the DFG.
\bibliographystyle{mn2e}
\bibliography{references}

\begin{thebibliography}{}

\bibitem[\protect\citeauthoryear{{Alongi}, {Bertelli}, {Bressan}, {Chiosi},
  {Fagotto}, {Greggio} \& {Nasi}}{{Alongi} et~al.}{1993}]{Alongi1993}
{Alongi} M.,  {Bertelli} G.,  {Bressan} A.,  {Chiosi} C.,  {Fagotto} F.,
  {Greggio} L.,    {Nasi} E.,  1993, \aaps, 97, 851

\bibitem[\protect\citeauthoryear{Behroozi, Wechsler \& Conroy}{Behroozi
  et~al.}{2012}]{Behroozi2012}
Behroozi P.~S.,  Wechsler R.~H.,    Conroy C.,  2012

\bibitem[\protect\citeauthoryear{{Bensby}, {Feltzing} \&
  {Lundstr{\"o}m}}{{Bensby} et~al.}{2003}]{Bensby2003}
{Bensby} T.,  {Feltzing} S.,    {Lundstr{\"o}m} I.,  2003, \aap, 410, 527

\bibitem[\protect\citeauthoryear{{Bensby}, {Feltzing}, {Lundstr{\"o}m} \&
  {Ilyin}}{{Bensby} et~al.}{2005}]{Bensby2005}
{Bensby} T.,  {Feltzing} S.,  {Lundstr{\"o}m} I.,    {Ilyin} I.,  2005, \aap,
  433, 185

\bibitem[\protect\citeauthoryear{{Bird}, {Kazantzidis}, {Weinberg}, {Guedes},
  {Callegari}, {Mayer} \& {Madau}}{{Bird} et~al.}{2013}]{Bird2013}
{Bird} J.~C.,  {Kazantzidis} S.,  {Weinberg} D.~H.,  {Guedes} J.,  {Callegari}
  S.,  {Mayer} L.,    {Madau} P.,  2013, ArXiv/1301.0620

\bibitem[\protect\citeauthoryear{{Bovy}, {Allende Prieto}, {Beers}, {Bizyaev},
  {da Costa}, {Cunha}, {Ebelke}, {Eisenstein}, {Frinchaboy}, {Garc{\'{\i}}a
  P{\'e}rez}, {Girardi}, {Hearty} \& {Hogg}}{{Bovy} et~al.}{2012}]{Bovy2012d}
{Bovy} J.,  {Allende Prieto} C.,  {Beers} T.~C.,  {Bizyaev} D.,  {da Costa}
  L.~N.,  {Cunha} K.,  {Ebelke} G.~L.,  {Eisenstein} D.~J.,  {Frinchaboy}
  P.~M.,  {Garc{\'{\i}}a P{\'e}rez} A.~E.,  {Girardi} L.,  {Hearty} F.~R.,
  {Hogg} D.~W.,  2012, \apj, 759, 131

\bibitem[\protect\citeauthoryear{{Bovy}, {Rix} \& {Hogg}}{{Bovy}
  et~al.}{2012}]{Bovy2012a}
{Bovy} J.,  {Rix} H.-W.,    {Hogg} D.~W.,  2012, \apj, 751, 131

\bibitem[\protect\citeauthoryear{{Bovy}, {Rix}, {Liu}, {Hogg}, {Beers} \&
  {Lee}}{{Bovy} et~al.}{2012}]{Bovy2012}
{Bovy} J.,  {Rix} H.-W.,  {Liu} C.,  {Hogg} D.~W.,  {Beers} T.~C.,    {Lee}
  Y.~S.,  2012, \apj, 753, 148

\bibitem[\protect\citeauthoryear{{Bressan}, {Fagotto}, {Bertelli} \&
  {Chiosi}}{{Bressan} et~al.}{1993}]{Bressan1993}
{Bressan} A.,  {Fagotto} F.,  {Bertelli} G.,    {Chiosi} C.,  1993, \aaps, 100,
  647

\bibitem[\protect\citeauthoryear{{Brook}, {Gibson}, {Martel} \&
  {Kawata}}{{Brook} et~al.}{2005}]{Brook2005}
{Brook} C.~B.,  {Gibson} B.~K.,  {Martel} H.,    {Kawata} D.,  2005, \apj, 630,
  298

\bibitem[\protect\citeauthoryear{{Brook}, {Kawata}, {Gibson} \&
  {Freeman}}{{Brook} et~al.}{2004}]{Brook2004}
{Brook} C.~B.,  {Kawata} D.,  {Gibson} B.~K.,    {Freeman} K.~C.,  2004, \apj,
  612, 894

\bibitem[\protect\citeauthoryear{{Brook}, {Kawata}, {Martel}, {Gibson} \&
  {Bailin}}{{Brook} et~al.}{2006}]{Brook2006}
{Brook} C.~B.,  {Kawata} D.,  {Martel} H.,  {Gibson} B.~K.,    {Bailin} J.,
  2006, \apj, 639, 126

\bibitem[\protect\citeauthoryear{{Brook}, {Stinson}, {Gibson}, {Ro{\v s}kar},
  {Wadsley} \& {Quinn}}{{Brook} et~al.}{012a}]{Brook2012}
{Brook} C.~B.,  {Stinson} G.,  {Gibson} B.~K.,  {Ro{\v s}kar} R.,  {Wadsley}
  J.,    {Quinn} T.,  2012a, \mnras, 419, 771

\bibitem[\protect\citeauthoryear{{Brook}, {Stinson}, {Gibson}, {Wadsley} \&
  {Quinn}}{{Brook} et~al.}{012b}]{Brook2012a}
{Brook} C.~B.,  {Stinson} G.,  {Gibson} B.~K.,  {Wadsley} J.,    {Quinn} T.,
  2012b, \mnras, 424, 1275

\bibitem[\protect\citeauthoryear{{Burstein}}{{Burstein}}{1979}]{Burstein1979}
{Burstein} D.,  1979, \apj, 234, 829

\bibitem[\protect\citeauthoryear{{Chabrier}}{{Chabrier}}{2003}]{Chabrier2003}
{Chabrier} G.,  2003, \pasp, 115, 763

\bibitem[\protect\citeauthoryear{{Chiba} \& {Beers}}{{Chiba} \&
  {Beers}}{2000}]{Chiba2000}
{Chiba} M.,  {Beers} T.~C.,  2000, AJ, 119, 2843

\bibitem[\protect\citeauthoryear{{Dalcanton} \& {Bernstein}}{{Dalcanton} \&
  {Bernstein}}{2000}]{Dalcanton2000}
{Dalcanton} J.~J.,  {Bernstein} R.~A.,  2000, \aj, 120, 203

\bibitem[\protect\citeauthoryear{{Dalcanton} \& {Bernstein}}{{Dalcanton} \&
  {Bernstein}}{2002}]{Dalcanton2002}
{Dalcanton} J.~J.,  {Bernstein} R.~A.,  2002, \aj, 124, 1328

\bibitem[\protect\citeauthoryear{{de Jong}, {Seth}, {Radburn-Smith}, {Bell},
  {Brown}, {Bullock}, {Courteau}, {Dalcanton}, {Ferguson}, {Goudfrooij},
  {Holfeltz}, {Holwerda}, {Purcell}, {Sick} \& {Zucker}}{{de Jong}
  et~al.}{2007}]{deJong2007}
{de Jong} R.~S.,  {Seth} A.~C.,  {Radburn-Smith} D.~J.,  {Bell} E.~F.,  {Brown}
  T.~M.,  {Bullock} J.~S.,  {Courteau} S.,  {Dalcanton} J.~J.,  {Ferguson}
  H.~C.,  {Goudfrooij} P.,  {Holfeltz} S.,  {Holwerda} B.~W.,  {Purcell} C.,
  {Sick} J.,    {Zucker} D.~B.,  2007, \apjl, 667, L49

\bibitem[\protect\citeauthoryear{{Foreman-Mackey}, {Hogg}, {Lang} \&
  {Goodman}}{{Foreman-Mackey} et~al.}{2012}]{Foreman-Mackey2012}
{Foreman-Mackey} D.,  {Hogg} D.~W.,  {Lang} D.,    {Goodman} J.,  2012, ArXiv
  e-prints

\bibitem[\protect\citeauthoryear{{Fuhrmann}}{{Fuhrmann}}{1998}]{Fuhrmann1998}
{Fuhrmann} K.,  1998, AAP, 338, 161

\bibitem[\protect\citeauthoryear{{Fuhrmann}}{{Fuhrmann}}{2008}]{Fuhrmann2008}
{Fuhrmann} K.,  2008, MNRAS, 384, 173

\bibitem[\protect\citeauthoryear{{Gibson}}{{Gibson}}{2002}]{Gibson2002}
{Gibson} B.~K.,  2002, in {Nomoto} K.,  {Truran} J.~W.,  eds, Cosmic Chemical
  Evolution Vol.~187 of IAU Symposium, {Stellar yields and chemical evolution}.
pp 159--163

\bibitem[\protect\citeauthoryear{{Gilmore} \& {Reid}}{{Gilmore} \&
  {Reid}}{1983}]{Gilmore1983}
{Gilmore} G.,  {Reid} N.,  1983, \mnras, 202, 1025

\bibitem[\protect\citeauthoryear{{Gilmore}, {Wyse} \& {Jones}}{{Gilmore}
  et~al.}{1995}]{Gilmore1995}
{Gilmore} G.,  {Wyse} R.~F.~G.,    {Jones} J.~B.,  1995, \aj, 109, 1095

\bibitem[\protect\citeauthoryear{{Hammer}, {Puech}, {Chemin}, {Flores} \&
  {Lehnert}}{{Hammer} et~al.}{2007}]{Hammer2007}
{Hammer} F.,  {Puech} M.,  {Chemin} L.,  {Flores} H.,    {Lehnert} M.~D.,
  2007, \apj, 662, 322

\bibitem[\protect\citeauthoryear{{Haywood}, {Di Matteo}, {Lehnert}, {Katz} \&
  {Gomez}}{{Haywood} et~al.}{2013}]{Haywood2013}
{Haywood} M.,  {Di Matteo} P.,  {Lehnert} M.,  {Katz} D.,    {Gomez} A.,  2013,
  ArXiv/1305.4663

\bibitem[\protect\citeauthoryear{{Ivezi{\'c}}, {Sesar}, {Juri{\'c}}, {Bond},
  {Dalcanton}, {Rockosi}, {Yanny}, {Newberg}, {Beers}, {Allende Prieto},
  {Wilhelm} \& {Lee}}{{Ivezi{\'c}} et~al.}{2008}]{Ivezic2008}
{Ivezi{\'c}} {\v Z}.,  {Sesar} B.,  {Juri{\'c}} M.,  {Bond} N.,  {Dalcanton}
  J.,  {Rockosi} C.~M.,  {Yanny} B.,  {Newberg} H.~J.,  {Beers} T.~C.,
  {Allende Prieto} C.,  {Wilhelm} R.,    {Lee} Y.~S.,  2008, \apj, 684, 287

\bibitem[\protect\citeauthoryear{{Juri{\'c}}, {Ivezi{\'c}}, {Brooks}, {Lupton},
  {Schlegel}, {Finkbeiner}, {Padmanabhan}, {Bond}, {Sesar} \&
  {Rockosi}}{{Juri{\'c}} et~al.}{2008}]{Juric2008}
{Juri{\'c}} M.,  {Ivezi{\'c}} {\v Z}.,  {Brooks} A.,  {Lupton} R.~H.,
  {Schlegel} D.,  {Finkbeiner} D.,  {Padmanabhan} N.,  {Bond} N.,  {Sesar} B.,
    {Rockosi} C.~M.,  2008, \apj, 673, 864

\bibitem[\protect\citeauthoryear{{Katz}}{{Katz}}{1992}]{Katz1992}
{Katz} N.,  1992, \apj, 391, 502

\bibitem[\protect\citeauthoryear{{Kazantzidis}, {Bullock}, {Zentner},
  {Kravtsov} \& {Moustakas}}{{Kazantzidis} et~al.}{2008}]{Kazantzidis2008}
{Kazantzidis} S.,  {Bullock} J.~S.,  {Zentner} A.~R.,  {Kravtsov} A.~V.,
  {Moustakas} L.~A.,  2008, \apj, 688, 254

\bibitem[\protect\citeauthoryear{{Kazantzidis}, {Zentner}, {Kravtsov},
  {Bullock} \& {Debattista}}{{Kazantzidis} et~al.}{2009}]{Kazantzidis2009}
{Kazantzidis} S.,  {Zentner} A.~R.,  {Kravtsov} A.~V.,  {Bullock} J.~S.,
  {Debattista} V.~P.,  2009, \apj, 700, 1896

\bibitem[\protect\citeauthoryear{{Kennicutt}}{{Kennicutt}}{1998}]{Kennicutt1998}
{Kennicutt} R.~C.,  1998, \apj, 498, 541+

\bibitem[\protect\citeauthoryear{{Klypin}, {Zhao} \& {Somerville}}{{Klypin}
  et~al.}{2002}]{Klypin2002}
{Klypin} A.,  {Zhao} H.,    {Somerville} R.~S.,  2002, \apj, 573, 597

\bibitem[\protect\citeauthoryear{{Kordopatis}, {Hill}, {Irwin}, {Gilmore},
  {Wyse}, {Tolstoy}, {de Laverny}, {Recio-Blanco}, {Battaglia} \&
  {Starkenburg}}{{Kordopatis} et~al.}{2013}]{Kordopatis2013}
{Kordopatis} G.,  {Hill} V.,  {Irwin} M.,  {Gilmore} G.,  {Wyse} R.~F.~G.,
  {Tolstoy} E.,  {de Laverny} P.,  {Recio-Blanco} A.,  {Battaglia} G.,
  {Starkenburg} E.,  2013, ArXiv e-prints

\bibitem[\protect\citeauthoryear{{Liu} \& {van de Ven}}{{Liu} \& {van de
  Ven}}{2012}]{Liu2012}
{Liu} C.,  {van de Ven} G.,  2012, ArXiv e-prints

\bibitem[\protect\citeauthoryear{{Loebman}, {Ro{\v s}kar}, {Debattista},
  {Ivezi{\'c}}, {Quinn} \& {Wadsley}}{{Loebman} et~al.}{2011}]{Loebman2011}
{Loebman} S.~R.,  {Ro{\v s}kar} R.,  {Debattista} V.~P.,  {Ivezi{\'c}} {\v Z}.,
   {Quinn} T.~R.,    {Wadsley} J.,  2011, \apj, 737, 8

\bibitem[\protect\citeauthoryear{{Lopez}, {Krumholz}, {Bolatto}, {Prochaska} \&
  {Ramirez-Ruiz}}{{Lopez} et~al.}{2011}]{Lopez2011}
{Lopez} L.~A.,  {Krumholz} M.~R.,  {Bolatto} A.~D.,  {Prochaska} J.~X.,
  {Ramirez-Ruiz} E.,  2011, \apj, 731, 91

\bibitem[\protect\citeauthoryear{{Macci{\`o}}, {Stinson}, {Brook}, {Wadsley},
  {Couchman}, {Shen}, {Gibson} \& {Quinn}}{{Macci{\`o}}
  et~al.}{2012}]{Maccio2012}
{Macci{\`o}} A.~V.,  {Stinson} G.,  {Brook} C.~B.,  {Wadsley} J.,  {Couchman}
  H.~M.~P.,  {Shen} S.,  {Gibson} B.~K.,    {Quinn} T.,  2012, \apjl, 744, L9

\bibitem[\protect\citeauthoryear{{Majewski}}{{Majewski}}{1993}]{Majewski1993}
{Majewski} S.~R.,  1993, \araa, 31, 575

\bibitem[\protect\citeauthoryear{{Minchev}, {Chiappini} \& {Martig}}{{Minchev}
  et~al.}{2012}]{Minchev2012}
{Minchev} I.,  {Chiappini} C.,    {Martig} M.,  2012, ArXiv e-prints

\bibitem[\protect\citeauthoryear{{Moster}, {Naab} \& {White}}{{Moster}
  et~al.}{2013}]{Moster2013}
{Moster} B.~P.,  {Naab} T.,    {White} S.~D.~M.,  2013, \mnras, 428, 3121

\bibitem[\protect\citeauthoryear{{Navarro}, {Abadi}, {Venn}, {Freeman} \&
  {Anguiano}}{{Navarro} et~al.}{2011}]{Navarro2011}
{Navarro} J.~F.,  {Abadi} M.~G.,  {Venn} K.~A.,  {Freeman} K.~C.,    {Anguiano}
  B.,  2011, \mnras, 412, 1203

\bibitem[\protect\citeauthoryear{{Nomoto}, {Thielemann} \& {Yokoi}}{{Nomoto}
  et~al.}{1984}]{Nomoto1984}
{Nomoto} K.,  {Thielemann} F.-K.,    {Yokoi} K.,  1984, \apj, 286, 644

\bibitem[\protect\citeauthoryear{{Nordstr{\"o}m}, {Mayor}, {Andersen},
  {Holmberg}, {Pont}, {J{\o}rgensen}, {Olsen}, {Udry} \&
  {Mowlavi}}{{Nordstr{\"o}m} et~al.}{2004}]{Nordstrom2004}
{Nordstr{\"o}m} B.,  {Mayor} M.,  {Andersen} J.,  {Holmberg} J.,  {Pont} F.,
  {J{\o}rgensen} B.~R.,  {Olsen} E.~H.,  {Udry} S.,    {Mowlavi} N.,  2004,
  \aap, 418, 989

\bibitem[\protect\citeauthoryear{{Norris}}{{Norris}}{1987}]{Norris1987}
{Norris} J.,  1987, \apjl, 314, L39

\bibitem[\protect\citeauthoryear{{Pellegrini}, {Baldwin}, {Brogan}, {Hanson},
  {Abel}, {Ferland}, {Nemala}, {Shaw} \& {Troland}}{{Pellegrini}
  et~al.}{2007}]{Pellegrini2007}
{Pellegrini} E.~W.,  {Baldwin} J.~A.,  {Brogan} C.~L.,  {Hanson} M.~M.,  {Abel}
  N.~P.,  {Ferland} G.~J.,  {Nemala} H.~B.,  {Shaw} G.,    {Troland} T.~H.,
  2007, \apj, 658, 1119

\bibitem[\protect\citeauthoryear{{Powell}}{{Powell}}{1964}]{Powell1964}
{Powell} M.~J.~D.,  1964, The Computer Journal, 7, 155

\bibitem[\protect\citeauthoryear{Press, Teukolsky, Vetterling \&
  Flannery}{Press et~al.}{2007}]{Press2007}
Press W.~H.,  Teukolsky S.~A.,  Vetterling W.~T.,    Flannery B.~P.,  2007,
  Numerical Recipes, The art of Scientific Computing, Third Edition, third edn.
Cambridge University Press

\bibitem[\protect\citeauthoryear{{Prochaska}, {Naumov}, {Carney}, {McWilliam}
  \& {Wolfe}}{{Prochaska} et~al.}{2000}]{Prochaska2000}
{Prochaska} J.~X.,  {Naumov} S.~O.,  {Carney} B.~W.,  {McWilliam} A.,
  {Wolfe} A.~M.,  2000, \aj, 120, 2513

\bibitem[\protect\citeauthoryear{{Reddy}, {Lambert} \& {Allende
  Prieto}}{{Reddy} et~al.}{2006}]{Reddy2006}
{Reddy} B.~E.,  {Lambert} D.~L.,    {Allende Prieto} C.,  2006, MNRAS, 367,
  1329

\bibitem[\protect\citeauthoryear{{Ruchti}, {Fulbright}, {Wyse}, {Navarro},
  {Parker}, {Reid}, {Seabroke}, {Siebert}, {Siviero}, {Steinmetz}, {Watson},
  {Williams} \& {Zwitter}}{{Ruchti} et~al.}{2010}]{Ruchti2010}
{Ruchti} G.~R.,  {Fulbright} J.~P.,  {Wyse} R.~F.~G.,  {Navarro} J.~F.,
  {Parker} Q.~A.,  {Reid} W.,  {Seabroke} G.~M.,  {Siebert} A.,  {Siviero} A.,
  {Steinmetz} M.,  {Watson} F.~G.,  {Williams} M.,    {Zwitter} T.,  2010,
  ApJL, 721, L92

\bibitem[\protect\citeauthoryear{{Sch{\"o}nrich} \& {Binney}}{{Sch{\"o}nrich}
  \& {Binney}}{2009}]{Schonrich2009}
{Sch{\"o}nrich} R.,  {Binney} J.,  2009, \mnras, 399, 1145

\bibitem[\protect\citeauthoryear{{Seth}, {Dalcanton} \& {de Jong}}{{Seth}
  et~al.}{2005}]{Seth2005}
{Seth} A.~C.,  {Dalcanton} J.~J.,    {de Jong} R.~S.,  2005, \aj, 130, 1574

\bibitem[\protect\citeauthoryear{{Spergel}, {Bean}, {Dor{\'e}}, {Nolta},
  {Bennett}, {Dunkley}, {Hinshaw}, {Jarosik} \& {Komatsu}}{{Spergel}
  et~al.}{2007}]{Spergel2007}
{Spergel} D.~N.,  {Bean} R.,  {Dor{\'e}} O.,  {Nolta} M.~R.,  {Bennett} C.~L.,
  {Dunkley} J.,  {Hinshaw} G.,  {Jarosik} N.,    {Komatsu} E.,  2007, \apjs,
  170, 377

\bibitem[\protect\citeauthoryear{{Stinson}, {Seth}, {Katz}, {Wadsley},
  {Governato} \& {Quinn}}{{Stinson} et~al.}{2006}]{Stinson2006}
{Stinson} G.,  {Seth} A.,  {Katz} N.,  {Wadsley} J.,  {Governato} F.,
  {Quinn} T.,  2006, \mnras, 373, 1074

\bibitem[\protect\citeauthoryear{{Stinson}, {Bailin}, {Couchman}, {Wadsley},
  {Shen}, {Nickerson}, {Brook} \& {Quinn}}{{Stinson}
  et~al.}{2010}]{Stinson2010}
{Stinson} G.~S.,  {Bailin} J.,  {Couchman} H.,  {Wadsley} J.,  {Shen} S.,
  {Nickerson} S.,  {Brook} C.,    {Quinn} T.,  2010, \mnras, 408, 812

\bibitem[\protect\citeauthoryear{{Stinson}, {Brook}, {Macci{\`o}}, {Wadsley},
  {Quinn} \& {Couchman}}{{Stinson} et~al.}{2013}]{Stinson2013}
{Stinson} G.~S.,  {Brook} C.,  {Macci{\`o}} A.~V.,  {Wadsley} J.,  {Quinn}
  T.~R.,    {Couchman} H.~M.~P.,  2013, \mnras, 428, 129

\bibitem[\protect\citeauthoryear{{Stinson}, {Brook}, {Prochaska}, {Hennawi},
  {Shen}, {Wadsley}, {Pontzen}, {Couchman}, {Quinn}, {Macci{\`o}} \&
  {Gibson}}{{Stinson} et~al.}{2012}]{Stinson2012a}
{Stinson} G.~S.,  {Brook} C.,  {Prochaska} J.~X.,  {Hennawi} J.,  {Shen} S.,
  {Wadsley} J.,  {Pontzen} A.,  {Couchman} H.~M.~P.,  {Quinn} T.,  {Macci{\`o}}
  A.~V.,    {Gibson} B.~K.,  2012, \mnras, p.~3506

\bibitem[\protect\citeauthoryear{{Thielemann}, {Nomoto} \&
  {Yokoi}}{{Thielemann} et~al.}{1986}]{Thielemann1986}
{Thielemann} F.-K.,  {Nomoto} K.,    {Yokoi} K.,  1986, \aap, 158, 17

\bibitem[\protect\citeauthoryear{{van der Kruit} \& {Searle}}{{van der Kruit}
  \& {Searle}}{1981}]{vanderKruit1981}
{van der Kruit} P.~C.,  {Searle} L.,  1981, \aap, 95, 105

\bibitem[\protect\citeauthoryear{{van der Kruit} \& {Searle}}{{van der Kruit}
  \& {Searle}}{1982}]{vanderKruit1982}
{van der Kruit} P.~C.,  {Searle} L.,  1982, \aap, 110, 61

\bibitem[\protect\citeauthoryear{{Villalobos}, {Kazantzidis} \&
  {Helmi}}{{Villalobos} et~al.}{2010}]{Villalobos2010}
{Villalobos} {\'A}.,  {Kazantzidis} S.,    {Helmi} A.,  2010, \apj, 718, 314

\bibitem[\protect\citeauthoryear{{Wadsley}, {Stadel} \& {Quinn}}{{Wadsley}
  et~al.}{2004}]{Wadsley2004}
{Wadsley} J.~W.,  {Stadel} J.,    {Quinn} T.,  2004, New Astronomy, 9, 137

\bibitem[\protect\citeauthoryear{{Woosley} \& {Weaver}}{{Woosley} \&
  {Weaver}}{1995}]{Woosley1995}
{Woosley} S.~E.,  {Weaver} T.~A.,  1995, \apjs, 101, 181

\bibitem[\protect\citeauthoryear{{Wyse}, {Gilmore}, {Norris}, {Wilkinson},
  {Kleyna}, {Koch}, {Evans} \& {Grebel}}{{Wyse} et~al.}{2006}]{Wyse2006}
{Wyse} R.~F.~G.,  {Gilmore} G.,  {Norris} J.~E.,  {Wilkinson} M.~I.,  {Kleyna}
  J.~T.,  {Koch} A.,  {Evans} N.~W.,    {Grebel} E.~K.,  2006, ApJL, 639, L13

\bibitem[\protect\citeauthoryear{Xue, Rix, Zhao, Fiorentin, Naab, Steinmetz,
  Van Den~Bosch, Beers, Lee, Bell et~al.,}{Xue et~al.}{2008}]{Xue2008}
Xue X.,  Rix H.,  Zhao G.,  Fiorentin P.,  Naab T.,  Steinmetz M.,  Van
  Den~Bosch F.,  Beers T.,  Lee Y.,  Bell E.,    et~al., 2008, The
  Astrophysical Journal, 684, 1143

\bibitem[\protect\citeauthoryear{{Yanny}, {Rockosi}, {Newberg}, {Knapp},
  {Adelman-McCarthy}, {Alcorn}, {Allam}, {Allende Prieto}, {An}, {Anderson},
  {Anderson}, {Bailer-Jones}, {Bastian}, {Beers}, {Bell} \&
  {Belokurov}}{{Yanny} et~al.}{2009}]{Yanny2009}
{Yanny} B.,  {Rockosi} C.,  {Newberg} H.~J.,  {Knapp} G.~R.,
  {Adelman-McCarthy} J.~K.,  {Alcorn} B.,  {Allam} S.,  {Allende Prieto} C.,
  {An} D.,  {Anderson} K.~S.~J.,  {Anderson} S.,  {Bailer-Jones} C.~A.~L.,
  {Bastian} S.,  {Beers} T.~C.,  {Bell} E.,    {Belokurov} V.,  2009, \aj, 137,
  4377

\bibitem[\protect\citeauthoryear{{Yoachim} \& {Dalcanton}}{{Yoachim} \&
  {Dalcanton}}{2006}]{Yoachim2006}
{Yoachim} P.,  {Dalcanton} J.~J.,  2006, \aj, 131, 226

\end{thebibliography}

\clearpage

\end{document}